\newif\iffigs\figstrue
\documentclass[a4paper,12pt]{article}
\usepackage{latexsym,amssymb,lscape,graphics}
\usepackage{graphicx}        
\usepackage{longtable}
\usepackage{multirow}
\usepackage{color}
\usepackage{slashed,epsfig}
\usepackage{amsfonts}

\newcommand{\e}{\textrm{e}}

\newtheorem{definizione}{Definition}[section]

\newcommand{\bd}{\begin{definizione}}
\newcommand{\ed}{\end{definizione}}

\def\IC{\relax\,\hbox{$\inbar\kern-.3em{\rm C}$}}
\def\IG{\relax\,\hbox{$\inbar\kern-.3em{\rm G}$}}
\def\IB{\relax{\rm I\kern-.18em B}}
\def\ID{\relax{\rm I\kern-.18em D}}
\def\IL{\relax{\rm I\kern-.18em L}}
\def\IF{\relax{\rm I\kern-.18em F}}
\def\IH{\relax{\rm I\kern-.18em H}}
\def\II{\relax{\rm I\kern-.17em I}}
\def\IN{\relax{\rm I\kern-.18em N}}
\def\IP{\relax{\rm I\kern-.18em P}}
\def\IQ{\relax\,\hbox{$\inbar\kern-.3em{\rm Q}$}}
\def\bfzero{\relax\,\hbox{$\inbar\kern-.3em{\rm 0}$}}
\def\IK{\relax{\rm I\kern-.18em K}}
\def\IG{\relax\,\hbox{$\inbar\kern-.3em{\rm G}$}}
 \font\cmss=cmss10 \font\cmsss=cmss10 at 7pt
\def\IR{\relax{\rm I\kern-.18em R}}
\def\ZZ{\relax\ifmmode\mathchoice
{\hbox{\cmss Z\kern-.4em Z}}{\hbox{\cmss Z\kern-.4em Z}}
{\lower.9pt\hbox{\cmsss Z\kern-.4em Z}} {\lower1.2pt\hbox{\cmsss
Z\kern-.4em Z}}\else{\cmss Z\kern-.4em Z}\fi}
\def\bfone{\relax{\rm 1\kern-.35em 1}}

\def\inbar{\vrule height1.5ex width.4pt depth0pt}
\def\bfzero{\relax{\rm I\kern-.18em 0}}
\def\bfone{\relax{\rm 1\kern-.35em 1}}

\def\o#1#2{{{#1}\over{#2}}}
\DeclareFontFamily{U}{rsf}{} \DeclareFontShape{U}{rsf}{m}{n}{
  <5> <6> rsfs5 <7> <8> <9> rsfs7 <10-> rsfs10}{}
\DeclareMathAlphabet\Scr{U}{rsf}{m}{n}

\def\e{\epsilon}

\newcommand{\ft}[2]{{\textstyle\frac{#1}{#2}}}

\def\1bar{1\hskip -.275cm -}
\def\2bar{2\hskip -.275cm -}
\def\3bar{3\hskip -.275cm -}

\newsavebox{\uuunit}
\sbox{\uuunit}
                 {\setlength{\unitlength}{0.825em}
                      \begin{picture}(0.6,0.7)
                                      \thinlines
                                      \put(0,0){\line(1,0){0.5}}
                                      \put(0.15,0){\line(0,1){0.7}}
                                      \put(0.35,0){\line(0,1){0.8}}
                                     \multiput(0.3,0.8)(-0.04,-0.02){10}{\rule{0.5pt}{0.5pt}}
                      \end {picture}}

\makeatletter \@addtoreset{equation}{section} \makeatother

\def\bfone{\relax{\rm 1\kern-.35em 1}}

\def\bfone{\relax{\rm 1\kern-.35em 1}}
\font\cmss=cmss10 \font\cmsss=cmss10 at 7pt

\newcommand{\so}{\mathfrak{so}}
\newcommand{\su}{\mathfrak{su}}
\newcommand{\usp}{\mathfrak{usp}}

\newcommand{\slal}{\mathfrak{sl}}

\def\bfone{\relax{\rm 1\kern-.35em 1}}
\def\inbar{\vrule height1.5ex width.4pt depth0pt}
\def\IC{\relax\,\hbox{$\inbar\kern-.3em{\rm C}$}}
\def\ID{\relax{\rm I\kern-.18em D}}
\def\IF{\relax{\rm I\kern-.18em F}}
\def\IH{\relax{\rm I\kern-.18em H}}
\def\II{\relax{\rm I\kern-.17em I}}
\def\IN{\relax{\rm I\kern-.18em N}}
\def\IP{\relax{\rm I\kern-.18em P}}
\def\IQ{\relax\,\hbox{$\inbar\kern-.3em{\rm Q}$}}
\def\IR{\relax{\rm I\kern-.18em R}}
\font\cmss=cmss10 \font\cmsss=cmss10 at 7pt
\def\ZZ{\relax\ifmmode\mathchoice
{\hbox{\cmss Z\kern-.4em Z}}{\hbox{\cmss Z\kern-.4em Z}} {\lower.9pt\hbox{\cmsss Z\kern-.4em Z}}
{\lower1.2pt\hbox{\cmsss Z\kern-.4em Z}}\else{\cmss Z\kern-.4em Z}\fi}

\def\e{\epsilon}

\def\bar{\overline}

\def\hat{\widehat}

\def\Coe#1.#2.{{#1\over #2}}

\def\coe#1.#2.{\relax{\textstyle {#1 \over #2}}\displaystyle}

\def\to{\rightarrow}
\def\notin{\hbox{{$\in$}\kern-.51em\hbox{/}}}

\def\IE{\relax{{\rm I\kern-.18em E}}}

\def\IGam{\relax{{\rm I}\kern-.18em \Gamma}}

\def\IA{\relax{\hbox{{\rm A}\kern-.82em {\rm A}}}}

\begin{document}
\begin{titlepage}
\begin{center}
\vskip 0.2cm
\vskip 0.2cm
{\LARGE \bf  Hyperinstantons, the Beltrami Equation,
\\[0.3cm]
and Triholomorphic Maps
}\\[0.5cm]
{ {\Large P.~Fr\'e}${}^{\; a,d,}$\footnote{Prof. Fr\'e is presently fulfilling the duties of Scientific
Counselor of the Italian Embassy in the Russian Federation, Denezhnij pereulok, 5, 121002 Moscow, Russia.
\emph{e-mail:} \quad {\small {\tt pietro.fre@esteri.it}}}, { {\Large P.A.~Grassi}${}^{\; b}$}, {\Large  and A.S.~Sorin}$^{\; c,d}$ \\[10pt]
\vspace{.2cm}
{${}^a$\sl\small Dipartimento di Fisica, Universit\`a di Torino\\INFN -- Sezione di Torino \\
via P. Giuria 1, \ 10125 Torino \ Italy\\}
\emph{e-mail:} \quad {\small {\tt fre@to.infn.it}}\\
\vspace{3pt}
{${}^b$\sl\small Dipartimento di Scienze e Innovazione Tecnologica,\\
Viale T. Michel 11, 15121 Alessandria, Italy
Universit\`a del Piemonte Orientale,\\
and INFN Sezione di Torino}\\
\emph{e-mail:} \quad {\small {\tt pietro.grassi@uniupo.it}}\\
\vspace{5pt}
{{\em $^{c}$\sl\small Bogoliubov Laboratory of Theoretical Physics and}}\\
{{\em Veksler and Baldin Laboratory of High Energy Physics}}\\
{{\em Joint Institute for Nuclear Research,}}\\
{\em 141980 Dubna, Moscow Region, Russia}~\quad\\
\emph{e-mail:}\quad {\small {\tt sorin@theor.jinr.ru}}\\
\vspace{3pt}
{{\em $^{d}$\sl\small  National Research Nuclear University MEPhI\\ (Moscow Engineering Physics Institute),}}\\
{\em Kashirskoye shosse 31, 115409 Moscow, Russia}~\quad\\
\quad \\
\quad \vspace{4pt}}
\vspace{10pt}
\begin{abstract}
\vspace{.3cm} \noindent We consider the Beltrami equation for hydrodynamics and we show that its solutions can be viewed as
instanton solutions of a more general system of equations. The latter are the equations of motion for an ${\cal N}=2$ sigma model
on 4-dimensional worldvolume (which is taken locally HyperK\"ahler) with a 4-dimensional HyperK\"ahler target space.  By means of
the 4D twisting procedure originally introduced by Witten for gauge theories and later generalized to 4D sigma-models by Anselmi
and Fr\'e, we show that the equations of motion describe triholomophic maps between the worldvolume and the target space.
Therefore, the classification of the solutions to the 3-dimensional Beltrami equation can be performed by counting the
triholomorphic maps. The counting is easily obtained by using several discrete symmetries. Finally, the similarity with
holomorphic maps for ${\cal N}=2$ sigma on Calabi-Yau space prompts us to reformulate the problem of the enumeration of
triholomorphic maps in terms of  a topological sigma model.
\end{abstract}
\end{center}
\end{titlepage}
\tableofcontents
\noindent {}
\newpage
\section{Introduction}
In a recent paper \cite{Fre:2015xaa}, two of us  proposed  what they described as a \textit{Sentimental Journey from Hydrodynamics to Supergravity}, namely a reinterpretation of the solutions of Beltrami equation as fluxes in 2-brane exact solutions of $D=7$ minimal supergravity \cite{PvNT},\cite{SalamSezgin},\cite{bershoffo1}. In the same jocose  spirit we can describe the present paper as
 a \textit{Sentimental Journey from Hydrodynamics to Hyperinstantons}, since we show hereby that the
 solutions  of Beltrami equation can be mapped into solutions of the \textit{triholomorphicity constraint} that defines the  hyperinstantons, namely the instanton solutions of  an $\mathcal{N}=2$ supersymmetric sigma model in D=4.
 \par
Hence  let us briefly recall the two sides of the correspondence which we plan to demonstrate. 
\par
The  first side is a simple first order differential equation written in the XIX century
by the great Italian Mathematician Eugenio Beltrami\cite{beltramus}:  an equation that bears his name and can be cast
in the following modern notation:
\begin{eqnarray}
    \star_g \mathrm{d} \mathbf{Y}_{[1]} &=& \, \mu \, \mathbf{Y}_{[1]}   \label{formaduale}
\end{eqnarray}
The unknown in this equation is a 1-form $\mathbf{Y}_{[1]} = \mathbf{Y}_{i} dx^i$. The symbol  $\star_g$ denotes the Hodge dual on a 3-dimensional manifold. Indeed  eq.(\ref{formaduale}) is an eigenvalue problem which makes sense only on three-manifolds $\mathcal{M}_3$. If $\mathcal{M}_3$ is compact, the spectrum of the $\star \,\mathrm{d}$ operator is discrete and encodes topological properties of the manifold. In particular if $\mathcal{M}_3$ is a flat torus $\mathrm{T^3}$, the whole spectrum of eigenvalues and eigenfunctions can be constructed with simple algorithms and it can be organized into
irreducible representations of a rich variety of crystallographic groups that were recently explored and classified by
two of us \cite{Fre:2015mla}. The hydrodynamical viewpoint on eq.(\ref{formaduale}) arises from the trivial
observation that a $1$-form $\mathbf{Y}_{[1]}$ is dual to a vector field $\vec{\mathbf{V}}$ and that any vector field
in three-dimensions can be interpreted as the velocity field of some fluid. This hydrodynamical interpretation of
eq.(\ref{formaduale}) is boosted by the existence of a very important theorem proved by V. Arnold
\cite{arnoldorussopapero}: \textit{on compact manifolds $\mathcal{M}_3$, streamlines of a steady flow have a chance of
displaying a chaotic behavior if and only if the one-form dual to the vector-field of the flow satisfies Beltrami
equation}. See on this point also \cite{balubbo},\cite{Childress}
\par
In order to approach the second side of the correspondence we aim to analyse, let us stress that (\ref{formaduale})
is the three dimensional counterpart of the (anti) self-duality condition for 4-dimensional instantons in a
gauge field theory. For a diagonal metric, it can be written as
\begin{equation}\label{tensoBeltra2}
    \frac{1}{2} \,  \epsilon_{ijk} \partial_j \mathbf{Y}_k \, = \, \mu  \, \mathbf{Y}_i\,.
\end{equation}
which implies $ \partial^i \mathbf{Y}_{i} =0$ if $\mu \neq 0$.
\par
If we embed a solution to (\ref{formaduale}) into a four dimensional manifold by adding a further component $\mathbf{Y}_{0}$ to the 1-form $\mathbf{Y}_{[1]}$  with a corresponding new coordinate $\mathrm{U}$, we can rewrite Beltrami equation as follows
\begin{eqnarray}
\label{newanto}
d \star_g \mathbf{q}_{[1]} =0\,, ~~~~~ (1 + \star_g) \, d  \mathbf{q}_{[1]} =0\
\end{eqnarray}
where $ \mathbf{q}_{[1]}$ is a 1-form on a four dimensional manifold and $\star_g$ is the Hodge dual on that manifold.
This set of equations were studied in \cite{Anselmi:1993wm}. Its authors showed that these
equations can be rephrased as {\it triholomorphic} maps from a flat 4-dimensional HyperK\"ahler manifold
to another HyperK\"alher.
In particular,
if we consider the components $\mathbf{q}_{s}$ (with $s =0,\dots,3$) as maps
$$q:{\cal M}_4\rightarrow {\cal N}_{4}$$
from a four dimensional flat HyperK\"ahler (worldvolume) manifold ${\cal M}_4$ to a (target) HyperK\"ahler manifold ${\cal N}_{4}$,
the equations (\ref{newanto}) are equivalent to
\begin{eqnarray}
\label{newanto2}
q^\star-J_x\circ q^\star\circ j_x=0. \label{afeq}
\end{eqnarray}
where $q^\star$ are the push-forward $T{\cal M}_4\rightarrow T{\cal N}_{4}$ and
$j^x$ and $J^x$ (with $x=1,..,3$) are the three complex structures of ${\cal M}_4$ and
of ${\cal N}_{4}$, respectively.
In the present paper we consider a $4 d$ topological $\sigma$-model as a theory of those maps.
The precise form of our topological field theory   is obtained from the topological twist of an
$\mathcal{N}=2$ $\sigma$-model. It can be easily generalized to $4m$-dimensional target space
${\cal N}_{4m}$ as discussed in the text.
\par
As we will recall in a later section, eq. (\ref{afeq}) arises uniquely from the topological twist of an $\mathcal{N}=2$
sigma-model in $D=4$ and it was proposed by Anselmi and Fr\'e in  \cite{Anselmi:1993wm} as the correct \textit{triholomorphicity} generalization of  the \textit{holomorphicity constraint}:
 \begin{equation}
   X^\star \, - \, J\circ X^\star\circ j=0. \label{holom}
 \end{equation}
 satisfied by a holomorphic map $X: \, \mathcal{M} \to \mathcal{N}$ from a complex manifold $\mathcal{M}$ to a complex manifold $\mathcal{N}$.
 Actually in \cite{Anselmi:1993wm} it was noted that the definition (\ref{afeq}) might be slightly generalized
since there is no uniqueness of the relative ordering of the three complex structures of the
two manifolds. It was proposed that eq.\ ({\ref{afeq}) might be substituted with the more
general condition
\begin{equation}
q^\star-\mathrm{O}^{xy}J_x\circ q^\star\circ j_y=0, \label{afeq4}
\end{equation}
where $\mathrm{O}^{xy}$ is an $\mathrm{SO(3)}$ matrix that can depend on the point. According to this weaker definition,
\textit{triholomorphic maps} are those maps $q$ for which there exists a $\mathrm{O}^{xy}$ such that (\ref{afeq4})
holds.
\par
Indeed the matrix $\mathrm{O}^{xy}$ is nothing else but a transition function of the
$\mathrm{SU(2)}$ bundle $\mathcal{SU}_\mathrm{I}$ associated with the triplet of K\"ahler $2$-forms that define the
HyperK\"ahler geometry. In the case of HyperK\"ahler geometry,
$\mathcal{SU}_\mathrm{I}$ is a flat bundle and the transition functions $\mathrm{O}^{xy}$ are necessarily constant.
\par
Once we have recognised that the triholomoprhic maps are in correspondence with the solutions of the
Beltrami equation, we construct those maps and we classify them according to the discrete group
studied in the previous papers of the subject. In particular, we construct several triholomorphic maps characterised
by the quantum number of the representation theory for the octhaedral group $\mathrm{{O}_{24}}$. In terms of those solutions,
we analyze the topological action, which reduces to the topological term, and we show how to disentangle the moduli of the
solution from the rest. In particular, we demonstrate the BRST symmetry of the action due to the boundary conditions on the
ghost fields evaluated on the triholomorphic maps.
\par
The paper is organized as follows. In sec. 2 we recall some basic ingrdients on HyperK\"alher geometry, notations and the two $\mathrm{SU(2)}$ bundles over the space needed to perform the twisting procedure. In sec. 2.3 we discuss flat HyperK\"alher
geometry and we introduce the fundamental notations we are using in the rest of the paper. In sec. 3, we discuss N=2 sigma model in four dimensions and its topological twist. On the latter part we insist to provide a complete and a self-contained discussion. In sec. 4 we discuss the relation among triholomorphic maps, Beltrami vector fields and hyperinstantons.
The explicit expressions of the solution to the triholomorphic equation is given and the classification of the solution
is performed by using the discrete groups discussed in the previous literature \cite{Fre:2015mla}. In sec. 5, the coupling
constants are discussed and in sec. 6 the moduli space is constructed. We discuss in detail the action, the functional integral and the boundary conditions for the ghost fields (twisted fermions). In the appendix we recall some properties of
triholomorphic hyperinstantons and other auxilliary material.
\par
\section{HyperK\"ahler Geometry}
\label{iperkallero}
Here we summarize the concepts and the definitions of HyperK\"ahler geometry
following \cite{Andrianopoli:1996cm}.
\par
A HyperK\"ahler manifold ${\cal HM}$ is a $4 m$-dimensional real manifold endowed with a metric $h$:
\begin{equation}
d s^2 = h_{u v} (q) d q^u \otimes d q^v   \quad ; \quad u,v=1,\dots,
4  m
\label{qmetrica}
\end{equation}
and three complex structures
\begin{equation}
(J^x) \,:~~ T({\cal HM}) \, \longrightarrow \, T({\cal HM}) \qquad
\quad
(x=1,2,3)
\end{equation}
that satisfy the quaternionic algebra
\begin{equation}
J^x J^y = - \delta^{xy} \, \bfone \,  +  \, \epsilon^{xyz} J^z
\label{quatalgebra}
\end{equation}
and respect to which the metric is hermitian:
\begin{equation}
\forall   \mbox{\bf X} ,\mbox{\bf Y}  \in   T{\cal HM}   \,: \quad
h \left( J^x \mbox{\bf X}, J^x \mbox{\bf Y} \right )   =
h \left( \mbox{\bf X}, \mbox{\bf Y} \right ) \quad \quad
  (x=1,2,3)
\label{hermit}
\end{equation}
From eq.~(\ref{hermit}) it follows that one can introduce a triplet of  2-forms
\begin{equation}
\begin{array}{ccccccc}
\mathbf{K}^x& = &K^x_{u v} d q^u \wedge d q^v & ; & K^x_{uv} &=&   h_{uw} (J^x)^w_v \cr
\end{array}
\label{iperforme}
\end{equation}
 The triplet $\mathbf{K}^x$ is named the {\it HyperK\"ahler} form: it is an $\mathrm{SU(2)}$ Lie--algebra valued
2--form.
\subsection{The flat $\mathrm{SU(2)_I}\,$-bundle and holonomy}
\label{su2Ibund} Let us  introduce a principal $\mathrm{SU(2)}$--bundle
\begin{equation}
{\mathcal{SU}}_\mathrm{I} \, \longrightarrow \, {\cal HM} \label{su2bundle}
\end{equation} and let $\omega^x$ denote a flat connection on such a bundle:
\begin{equation}
 d \omega^x +
{1\over 2} \epsilon^{x y z} \omega^y \wedge \omega^z \, = \, 0
\label{su2curv}
\end{equation}
The  definition of  a HyperK\"ahler manifold requires that  the HyperK\"ahler 2--form $\mathbf{K}^x$
should be covariantly closed
with respect to the flat  connection $\omega^x$:
\begin{equation}
\nabla \mathbf{K}^x \equiv \mathrm{d} \mathbf{K}^x + \epsilon^{x y z} \omega^y \wedge \mathbf{K}^z    \, = \, 0
\label{closkform}
\end{equation}
In any local patch of the bundle the flat-connection $\omega^y$ can be reduced to zero and in that patch the
HyperK\"ahler form is closed $d \mathbf{K}^x \, = \, 0$. The structural group of the above mentioned flat bundle of which the
HyperK\"ahler $2$-forms constitute a section is named $\mathrm{SU(2)_I}$. It plays an important role in supersymmetry
since in the construction of a $\mathcal{N}=2$ sigma-model it is identified with the $\mathrm{SU(2)}$-automorphism
group of the supersymmetry algebra.
\par
As a consequence of the above structure the HyperK\"ahler manifold ${\cal HM}$ has a
holonomy group of the following type:
\begin{eqnarray}
{\rm Hol}({\cal HM})&=& \bfone \otimes {\cal H} \quad ; \quad {\cal H} \,  \subset \,  \mathrm{USp} (2m)
\label{olonomia}
\end{eqnarray}
Introducing flat indices $\{A,B,C= 1,2\}, \{{\cal I},{\cal J},{\cal K} = 1,.., 2m\}$  that run, respectively, in the
fundamental representations of $\mathrm{\mathrm{SU(2)}_I}$ and $\mathrm{USp(2m)}$ (we denote by
$\su(2)_\mathrm{I},\usp(\mathrm{2m})$ the corresponding Lie algebras), we can find a complex vielbein
1-form
\begin{equation}
{\cal U}^{A{\cal I}} = {\cal U}^{A{\cal I}}_u (q) d q^u
\label{quatvielbein}
\end{equation}
such that
\begin{equation}
h_{uv} = {\cal U}^{A{\cal I}}_u {\cal U}^{B{\cal J}}_v
\IC_{{\cal I}{\cal J}}\epsilon_{AB}
\label{quatmet}
\end{equation}
where $\IC_{{\cal I} {\cal J}} = - \IC_{{\cal J} {\cal I}}$ and $\epsilon_{AB} = - \epsilon_{BA}$ are, respectively,
the flat $\mathrm{USp(2m)}$ and $\mathrm{USp(2)} \sim \mathrm{SU(2)_I}$ invariant metrics. The vielbein ${\cal
U}^{A{\cal I}}$ is covariantly closed with respect to the flat $\su(2)_\mathrm{I}\,$-connection $\omega^z$ and to some
$\usp(\mathrm{2m})$-Lie Algebra valued connection $\Delta^{{\cal I}{\cal J}} = \Delta^{{\cal J} {\cal I}}$:
\begin{eqnarray}
\nabla {\cal U}^{A{\cal I}}& \equiv & d{\cal U}^{A{\cal I}}
+{i\over 2} \omega^x (\epsilon \sigma_x\epsilon^{-1})^A_{\phantom{A}B}
\wedge{\cal U}^{B{\cal I}} + \Delta^{{\cal I}{\cal J}} \wedge {\cal U}^{A{\cal K}} \IC_{{\cal J}{\cal K}}
=0
\label{quattorsion}
\end{eqnarray}
\noindent
where $(\sigma^x)_A^{\phantom{A}B}$ are the standard Pauli matrices.
Furthermore ${ \cal U}^{A{\cal I}}$ satisfies  the reality condition:
\begin{equation}
{\cal U}_{A{\cal I}} \equiv ({\cal U}^{A{\cal I}})^* = \epsilon_{AB}
\IC_{{\cal I}{\cal J}} {\cal U}^{B{\cal J}}
\label{quatreality}
\end{equation}
We have also the inverse vielbein ${\cal U}^u_{A{\cal I}}$ defined by the
equation
\begin{equation}
{\cal U}^u_{A{\cal I}} {\cal U}^{A{\cal I}}_v = \delta^u_v
\label{2.64}
\end{equation}
Flattening a pair of indices of the Riemann
tensor ${\cal R}^{uv}_{\phantom{uv}{ts}}$
we obtain
\begin{equation}
{\cal R}^{uv}_{\phantom{uv}{ts}} {\cal U}^{{\cal I} A}_u {\cal U}^{{\cal J} B}_v =
 \IR^{{\cal I}{\cal J}}_{ts}\epsilon^{AB}
\label{2.65}
\end{equation}
\noindent where $\IR^{{\cal I}{\cal J}}_{ts}$ is the field strength of the $\usp(\mathrm{2m})$-connection:
\begin{equation}
\mathrm{d} \Delta^{{\cal I}{\cal J}} + \Delta^{{\cal I} {\cal K}} \wedge \Delta^{{\cal L} {\cal J}} \IC_{{\cal K} {\cal
L}} \equiv \IR^{{\cal I}{\cal J}} = \IR^{{\cal I} {\cal J}}_{ts} dq^t \wedge dq^s \label{2.66}
\end{equation}
Eq. ~(\ref{2.65}) is the explicit statement that the Levi Civita connection
associated with the metric $h$ has a holonomy group contained in
$\mathbf{1 }\otimes \mathrm{USp(2m)}$.
Finally we have the following relation between the HyperK\"ahler form and the complex vielbein
\begin{equation}
\mathbf{K}^x =\,-{\rm i}\, \ft 12 \, (\sigma _x)_A^{\phantom {A}C}\, {\cal U}^{{\cal I} A} \wedge {\cal U}_{{\cal I} C}
\label{2.69}
\end{equation}
From  equation (\ref{2.69}) one easily retrieves the following useful identity:
\begin{equation}\label{usefulID}
    {\rm i} \, \sigma_{AB}^x \, \mathcal{U}^{B \mathcal{I}}_u \, = \, \epsilon_{AB} \, J^{x |v}_{u}\,
    \mathcal{U}^{B \mathcal{I}}_v
\end{equation}
The above equation has the following clearcut geometrical interpretation. The action of the quaternionic tensors on the
tangent bundle can always be compensated by a transformation in the $\mathrm{SU(2)_I}\,$-fiber of the
${SU}_\mathrm{I}$-bundle.
\subsection{The  $\mathrm{SU(2)_Q}\,$-bundle and reduced holonomy}
\label{su2Qbund} Besides ${\mathcal{SU}_I}$ there is a second $\mathrm{SU(2)}$ principal bundle defined by
HyperK\"ahler geometry, which is not necessarily flat. This bundle naturally arises from the simple algebraic
consideration that there exists the following universal subalgebra:
\begin{equation}\label{qsubalg}
    \su(2)_\mathrm{Q}\oplus \so(\mathrm{m}) \, \subset \, \usp(\mathrm{2m})
\end{equation}
of the compact symplectic algebra. The procedure of the topological twist is well defined for those HyperK\"ahler
target manifolds $\mathcal{N}_{4m}$ where the holonomy is further reduced\footnote{This reduction of holonomy was
implicitly assumed in the construction of \cite{Anselmi:1993wm} but was not explicitly spelled out in that paper. This is very  important, since it provides a clear geometrical interpretation of the topological twist
as a basis for the introduction of hyperinstanton equations.} with respect to eq.(\ref{olonomia}), namely where we
have:
\begin{eqnarray}
{\rm Hol}(\mathcal{N}_{4m})&=& \bfone \otimes {\cal H} \quad ; \quad {\cal H} \,  \subset \,  \su(2)_\mathrm{Q}\oplus
\so(\mathrm{m}) \label{olonomiaRest}
\end{eqnarray}
When the condition (\ref{olonomiaRest}) is realized, we can use a refined index notation. The indices
$\mathcal{I},\mathcal{J},\mathcal{K},\dots$ taking $2m$-values can be substituted by  pairs of indices
$(\dot{A},k),(\dot{B},h),(\dot{C},\ell),\dots$ where $\dot{A},\dot{B},\dot{C},\dots \, =\,\dot{1},\dot{2}$ span the
fundamental representation of $\su(2)_\mathrm{Q}$, while $k,h,\ell,\dots \, = \, 1,2,3,\dots$ span the fundamental
representation of $\so(\mathrm{m})$. In this way the complex vielbein can be rewritten as follows:
\begin{equation}\label{colonnello1}
    \mathcal{U}^{A \mathcal{I}}_u \, = \, \mathcal{U}^{\phantom{u|a}A}_{u|a\phantom{A}\dot{B}}\quad ; \quad
     \mathcal{U}_{u|A \mathcal{I}} \, = \, \mathcal{U}_{u|a A}^{\phantom{u|a A} \dot{B}}
\end{equation}
and the symplectic metric $\mathbb{C}_{\mathcal{I}\mathcal{J}}$ becomes:
\begin{equation}\label{kuriashy}
    \mathbb{C}_{\mathcal{I}\mathcal{J}} \, = \, \epsilon^{\dot{A}\dot{B}}\, \delta_{hk} \quad ;
    \quad \mathbb{C}^{\mathcal{I}\mathcal{J}} \, = \, \epsilon_{\dot{A}\dot{B}}\, \delta_{hk}
\end{equation}
Equation (\ref{colonnello1}) admits a further rewriting which is very important at the level of the topological twist.
Indeed, using the quaternionic basis matrices (\ref{quatbas}) discussed in next section we can set:
\begin{eqnarray}
\mathcal{U}^{\phantom{k}A}_{k\phantom{A}\dot{B}}\, \equiv \,  \mathcal{U}^{\phantom{u|k}A}_{u|k\phantom{A}\dot{B}}\,
\mathrm{d}q^u &=& \left(e_\mu
  \right)^{A}_{\phantom{A}\dot{B}}\, E^\mu_{u|k} \,
\mathrm{d}q^u \, \equiv \, \left(e_\mu
  \right)^{A}_{\phantom{A}\dot{B}}\, \mathbf{E}^\mu_{k} \,\nonumber\\
\mathcal{U}_{k A}^{\phantom{a A} \dot{B}}\,\equiv \,  \mathcal{U}_{u|k A}^{\phantom{u|a A} \dot{B}}\, \mathrm{d}q^u &=&
\left(e^\dagger_\mu
  \right)_{A}^{\phantom{A}\dot{B}}\,E^\mu_{u|k}\, \mathrm{d}q^u \, \equiv \,\left(e^\dagger_\mu
  \right)_{A}^{\phantom{A}\dot{B}}\,\mathbf{E}^\mu_{k}\label{guliatori}
\end{eqnarray}
where:
\begin{equation}\label{forniciarino}
    \mathbf{E}^\mu_{k} \, = \,E^\mu_{u|k}\, \mathrm{d}q^u
\end{equation}
is a real vielbein $1$-form transforming in the bi--fundamental representation of ${\so(4)}_{\mathrm{IQ}}\oplus
\so(\mathrm{m})$, the algebra ${\so(4)}_{\mathrm{IQ}}$ being defined as follows:
\begin{equation}\label{meconsuelo}
    {\so(4)}_{\mathrm{IQ}}\, \equiv \, \su(2)_\mathrm{I}\oplus \su(2)_\mathrm{Q}
\end{equation}
As we discuss below in section \ref{topocurvo}, the geometrical basis of the topological twist is the identification
of the group $\so(4)_{\mathrm{IQ}}$ with the euclidianized Lorentz group $\so(4)_{\mathrm{Lorentz}}$. In order to be
able to do such an identification the existence of $\so(4)_{\mathrm{IQ}}$ is obviously necessary and the condition for
its existence is the reduced holonomy (\ref{olonomiaRest}) of the target HyperK\"ahler manifold $\mathcal{N}_{4m}$.
\par
One further identity which will be quite useful for the topological twist is the following one:
\begin{equation}\label{iddaEst}
   \ft 12 \, \mbox{Tr}\left[ e^\dagger_\mu \, e_\nu \, e_x \right]\, = \,J^{+|x}_{\mu\nu} \, \quad ; \quad x\,=\,
   1,2,3 \quad ; \quad \mu,\nu \, =\, 0,1,2,3
\end{equation}
In the above equation, $J^{+|x}_{\mu\nu}$ denotes the self-dual matrices discussed in next
section.
\subsubsection{Gamma matrices}
\label{gammamatti} Before proceeding further it is also convenient to fix a well adapted basis of $\so(4)$-gamma
matrices that we will use in dealing with the topological twist of the $\mathcal{N}=2$ sigma-model. We utilize a chiral
basis in which the matrix $\gamma_5$ is block diagonal. We set:
\begin{eqnarray}
  \gamma_0 &=& \sigma_1 \otimes \mathbf{1}_{2\times 2} \nonumber\\
  \gamma_x &=& \sigma_2 \otimes \sigma_x \quad (x=,1,2,3)\nonumber \\
  \gamma_5 &=& \sigma_3 \otimes \mathbf{1}_{2\times 2} \label{gammatine}
\end{eqnarray}
In this way we obtain that:
\begin{equation}\label{birrafresca}
    \gamma_\mu \, = \, \left(\begin{array}{c|c}
                               0 & e^\dagger_\mu \\
                               \hline
                               e_\mu & 0
                             \end{array}
     \right) \quad \mu\, = \, 0,1,2,3
\end{equation}
\subsection{Flat HyperK\"ahler Geometry  in $d=4$}
\label{flatkallero} In the present subsection
we analyse  the specific form of a flat four dimensional HyperK\"ahler manifold $\mathcal{HK}_{4}$. We begin first by
discussing its universal local geometry, then we discuss the choices of its global structure that are relevant to our
purposes.
\par
Considering the standard Pauli matrices:
\begin{equation}\label{sigPau}
 \sigma_1 \, = \,  \left(
\begin{array}{ll}
 0 & 1 \\
 1 & 0
\end{array}
\right) \quad; \quad \sigma_2 \, = \,\left(
\begin{array}{ll}
 0 & -i \\
 i & 0
\end{array}
\right)\quad; \quad \sigma_3 \, = \,\left(
\begin{array}{ll}
 1 & 0 \\
 0 & -1
\end{array}
\right)
\end{equation}
let us define the quaternionic basis as follows:
\begin{equation}\label{quatbas}
   e_0 \, = \, \left(\begin{array}{cc}
                       1 & 0 \\
                       0 & 1
                     \end{array}
   \right) \quad ; \quad e_x \, = \, {\rm i} \, \sigma_x \quad (x=1,2,3)
\end{equation}
and let us introduce the 't Hooft matrices, namely two sets of $4\times4$ antisymmetric matrices that represent  the
quaternionic algebra (\ref{quatalgebra}) and are respectively self--dual and antiself--dual:
\begin{equation}\label{selfdualJ}
    \begin{array}{ccccccccc}
       J^{+|1} & = & \left(
\begin{array}{llll}
 0 & -1 & 0 & 0 \\
 1 & 0 & 0 & 0 \\
 0 & 0 & 0 & -1 \\
 0 & 0 & 1 & 0
\end{array}
\right), & J^{+|2} & = & \left(
\begin{array}{llll}
 0 & 0 & -1 & 0 \\
 0 & 0 & 0 & 1 \\
 1 & 0 & 0 & 0 \\
 0 & -1 & 0 & 0
\end{array}
\right), & J^{+|3} & = & \left(
\begin{array}{llll}
 0 & 0 & 0 & -1 \\
 0 & 0 & -1 & 0 \\
 0 & 1 & 0 & 0 \\
 1 & 0 & 0 & 0
\end{array}
\right)
     \end{array}
\end{equation}
\begin{equation}\label{antiselfdualJ}
    \begin{array}{ccccccccc}
       J^{-|1} & = &  \left(
\begin{array}{llll}
 0 & -1 & 0 & 0 \\
 1 & 0 & 0 & 0 \\
 0 & 0 & 0 & 1 \\
 0 & 0 & -1 & 0
\end{array}
\right), & J^{-|2} & = & \left(
\begin{array}{llll}
 0 & 0 & 1 & 0 \\
 0 & 0 & 0 & 1 \\
 -1 & 0 & 0 & 0 \\
 0 & -1 & 0 & 0
\end{array}
\right), & J^{-|3} & = &  \left(
\begin{array}{llll}
 0 & 0 & 0 & -1 \\
 0 & 0 & 1 & 0 \\
 0 & -1 & 0 & 0 \\
 1 & 0 & 0 & 0
\end{array}
\right)
\end{array}
\end{equation}
\begin{equation}\label{selfata}
    J^{\pm|x}_{ab} \, = \,\pm \,\ft 12 \, \epsilon_{abcd}\, J^{\pm|x}_{cd}
\end{equation}
The matrices $J^{\pm|x} $ satisfy the $\su(2)$ Lie algebra and commute among themselves:
\begin{equation}\label{carolinaMoya}
   \left[ J^{\pm|x}\, , \, J^{\pm|y}\right ] \, = \, 2\, \epsilon_{xyz} \,J^{\pm|z} \quad ;
   \quad \left[ J^{\pm|x}\, , \, J^{\mp|y}\right ] \, = \,0
\end{equation}
Together the six generators $J^{\pm|x} $ span the $\so(4)$ Lie algebra and correspond to its decomposition:
\begin{equation}\label{skopettus}
    \so(4) \, = \,\su(2)_\mathrm{L} \oplus \su(2)_\mathrm{R}
\end{equation}
Given these conventions we name
\begin{equation}
q^u\, = \, \left\{\mathrm{U},\mathrm{X},\mathrm{Y},\mathrm{Z}\right\} \label{kantarellus}
\end{equation}
the four real coordinates of the flat HyperK\"ahler manifold and we introduce the quaternionic coordinate $\mathcal{Q}$
and the complex vielbein as it follows:
\begin{eqnarray}
  \mathcal{Q} &\equiv & \mathrm{U} \, e_0 \, + \, \mathrm{X} \, \e_1 \, + \, \mathrm{Y} \, e_2 \,
  + \, \mathrm{Z} \, e_3 \, = \, \left(
\begin{array}{ll}
 \mathrm{U}+i \mathrm{Z} & i \mathrm{X}+\mathrm{Y} \\
 i \mathrm{X}-\mathrm{Y} & \mathrm{U}-i \mathrm{Z}
\end{array}
\right)\nonumber\\
\mathcal{U} \, \equiv \, \mathrm{d}\mathcal{Q} \, = \,  \mathcal{U}^{A\mathcal{I}} &=& \mathrm{dU} \, e_0 \, +
   \, \mathrm{dX} \, \e_1 \, + \, \mathrm{dY} \, e_2 \,+ \, \mathrm{dZ} \, e_3 \, = \,\left(
\begin{array}{ll}
 \mathrm{dU}+i \mathrm{dZ} & i \mathrm{dX}+\mathrm{dY} \\
 i \mathrm{dX}-\mathrm{dY} & \mathrm{dU}-i \mathrm{dZ}
\end{array}
\right)\label{totocotugno}
\end{eqnarray}
Then according to formula (\ref{2.69}) we get:
\begin{equation}\label{consollo}
    \mathbf{K}^x \, = \, - \, {\rm i}\ft 12  \, \mbox{Tr} \, \left[ \mathcal{U}^T \wedge \sigma^x \,
    \mathcal{U}^\star \right]\, = \, \mathbf{K}^x_{[-]} \, \equiv \,
    J^{-|x}_{uv} \, dq^u \wedge dq^v
\end{equation}
By explicit substitution we obtain:
\begin{equation}\label{esplizzoHypK}
    \mathbf{K}^1_{[-]} \, = \, 2 \mathrm{dY}\wedge \mathrm{dZ}-2 \mathrm{dU}\wedge
   \mathrm{dX} \quad ; \quad  \mathbf{K}^2_{[-]} \, = \, 2 \mathrm{dU}\wedge \mathrm{dY}+2 \mathrm{dX}\wedge
   \mathrm{dZ} \quad ; \quad  \mathbf{K}^3_{[-]} \, = \, 2 \mathrm{dX}\wedge \mathrm{dY}-2 \mathrm{dU}\wedge
   \mathrm{dZ}
\end{equation}
On the other hand the metric $h$  has the following explicit appearance:
\begin{equation}\label{metricolana}
    ds^2 \, = \, h_{uv} \, dq^u \otimes dq^v \, = \, \mathrm{dU}^2 \, + \,\mathrm{dX}^2\, + \,
    \mathrm{dY}^2 \, + \,\mathrm{dZ}^2
\end{equation}
In addition to the HyperK\"ahler $2$-forms mentioned in (\ref{esplizzoHypK}) we have three self-dual $2$-forms:
\begin{equation}\label{esplizzoHypK}
    \mathbf{K}^1_{[+]} \, = \, 2 \mathrm{dY}\wedge \mathrm{dZ}+2 \mathrm{dU}\wedge
   \mathrm{dX} \quad ; \quad  \mathbf{K}^2_{[+]} \, = \, - \,2 \mathrm{dU}\wedge \mathrm{dY}+2 \mathrm{dX}\wedge
   \mathrm{dZ} \quad ; \quad  \mathbf{K}^3_{[+]} \, = \,-\, 2 \mathrm{dX}\wedge \mathrm{dY}-2 \mathrm{dU}\wedge
   \mathrm{dZ}
\end{equation}
Altogether we can arrange the 6 two-forms $\mathbf{K}^x_{[\mp]}$ into a $6$-vector:
\begin{equation}\label{arappo}
    \alpha^I =
    \left\{\underbrace{\mathbf{K}^1_{[-]},\mathbf{K}^2_{[-]},\mathbf{K}^3_{[-]}}_{I=1,2,3},
    \underbrace{\mathbf{K}^1_{[+]},\mathbf{K}^2_{[+]},\mathbf{K}^3_{[+]}}_{I=\dot{1},\dot{2},\dot{3}}\right\}
\end{equation}
with the properties:
\begin{eqnarray}
  \alpha^I \wedge \alpha^J &=&  8 \, \eta^{IJ} \, \mbox{Vol} \nonumber\\
  \mbox{Vol} &=& \mathrm{dU}\wedge \mathrm{dX}\wedge \mathrm{dY} \wedge \mathrm{dZ} \label{intersecuto1}
\end{eqnarray}
where the \textit{intersection matrix} is:
\begin{equation}\label{eta33primus}
    \eta^{IJ} \, = \, \left(\begin{array}{c|c}
                             -\, \mathbf{1}_{3\times3} & \mathbf{0}_{3\times 3} \\
                              \hline
                              \mathbf{0}_{3\times 3} & \mathbf{1}_{3\times3}
                            \end{array}
     \right)
\end{equation}
The group of linear transformations in the basis of $2$-forms $\alpha^I$ which preserves the intersection matrix and hence the
cohomology lattice is $\so(3,3)$. This will be relevant in the discussion of the lagrangian moduli.
\par
 We also need the boundary conditions to define the global topological structure of $\mathcal{HK}_4$. In the sequel
we consider two cases
\begin{equation}\label{romoaldino}
    \mathcal{HK}_4 \, = \, \mathcal{M}_4 \, \equiv \,\mathbb{R}_+ \times \mathrm{T^3} \quad ; \quad
    \mathcal{HK}_4 \, = \,\mathcal{N}_4 \, \equiv \, \mathbb{R}^4
\end{equation}
the first choice being utilized as the base manifold, the second as the target manifold in the topological
sigma-model.   Both  the manifold $\mathcal{M}_4$ and $\mathcal{N}_4$ are non compact, yet there is a
fundamental difference, the base manifold has  a boundary $\partial \mathcal{M}_4 \, \simeq \, \mathrm{T^3}$
corresponding to $\mathrm{U}=0$, while $\mathcal{N}_4$ has no boundary.

\par
In practice we obtain the first manifold $\mathcal{M}_4$ by means of the following conditions:
\begin{eqnarray}
    \mathrm{U} & \in & [0,\infty)\label{Urange1}\\
\left\{\mathrm{X,Y,Z}\right\} & \simeq & \left\{\mathrm{X,Y,Z}\right\} + \{n,m,r\}\quad ; \quad n,m,r\in
\mathbb{Z}\label{XYZ} \label{ciccipasticci1}
\end{eqnarray}
The second manifold $\mathcal{N}_4$ is obtained in the usual way assuming:
\begin{eqnarray}\label{ciccipasticci}
    \mathrm{q}^u & \in & (-\infty,\infty) \quad ; \quad u=0,1,2,3 \label{Urange2}
\end{eqnarray}
where, in order to distinguish them from the coordinates of the base manifold $\mathcal{M}_4$, we have renamed the
coordinates of the target manifold $\mathcal{N}_4$ as it follows:
\begin{equation}\label{rinuncino}
   \mathrm{ U \mapsto q^0 \quad ; \quad X \mapsto q^1 \quad ; \quad \quad Y \mapsto q^2; \quad \quad Z \mapsto q^3}
\end{equation}


\section{The $\mathcal{N}=2$ sigma-model in $D=4$ and its topological twist}
\label{topocurvo}
In this section, following once again \cite{Andrianopoli:1996cm} and \cite{Anselmi:1993wm} we present the
general form of a rigid $\mathcal{N}=2$ supersymmetric sigma-model in $D=4$ and we perform its topological twist according to the $4D$-algorithm developed by Anselmi and Fr\'e in \cite{Anselmi:1992tj},\cite{Anselmi:1992tz},\cite{Anselmi:1994bu}.
\par
Given the base flat manifold $\mathcal{M}_4$, which has always a local HyperK\"ahler structure, supersymmetry requires
that the target manifold should be a HyperK\"ahler manifold $\mathcal{N}_{4m}$ of real dimension $\mathrm{4m}$. The
field content of this theory is provided by the $4m$ scalar fields $q^u$ and by two chiral spin one-half fields, the
hyperini $\gamma_5 \, \zeta_\mathcal{I}\, = \, \zeta_\mathcal{I}$ and $\gamma_5 \,\zeta^\mathcal{I} \, = \, -
\,\zeta^\mathcal{I}$ that transform respectively in the $\mathbf{2m}$ and $\overline{\mathbf{2m}}$ representations of
the holonomy group $\mathrm{USp(2m)}$.
\par
Utilizing the geometric structures that we have introduced in section \ref{iperkallero}, the
 action of the rigid $\mathcal{N}=2$ sigma model can be written as follows (see
\cite{Andrianopoli:1996cm}):
\begin{eqnarray}
\mathcal{A}^{\mathcal{N}=2} = \int d^4x \Big[
h_{uv}
\partial^\mu q^u \, \partial_\mu  q^v
&-&   {\rm i}
\Bigl ( {\bar \zeta}^{{\cal I}} \, \gamma^\mu \, \partial_\mu \zeta_{{\cal I}} + {\bar \zeta}_{{\cal I}}
\gamma^\mu \, \partial_\mu
\zeta^{{\cal I}}\Bigr )
\nonumber\\
&  + & \, {\o{1}{2}} \, \IR^{\cal I}_{\phantom{{\cal I}}{\cal J}\vert ts}\, {\cal U}^t_{A{\cal K}} \,{\cal
U}^s_{B{\cal L}} \, \varepsilon^{AB} \, \IC^{{\cal L}{\cal P}} \, {\bar \zeta}_{\cal I} \, \zeta_{\cal P} \,   {\bar
\zeta}^{\cal J} \, \zeta^{\cal K} \Big]
\label{kinerige}
\end{eqnarray}
the four-Fermi interactions being dictated  by the $\usp(2m)$ curvature $\IR^{\cal I}_{\phantom{{\cal I}}{\cal J}\vert
ts}$ of the HyperK\"ahler target manifold.
\par
The supersymmetry transformations with respect to which the action (\ref{kinerige}) is invariant have the following
form (see \cite{Andrianopoli:1996cm}):
\begin{eqnarray}
{\cal U}^{{\cal I} A}_u (q) \, \delta q^u & = & \varepsilon^{AB} \, \IC^{{\cal I}{\cal J}} \, {\bar c_B} \zeta_{\cal J}
+ {\bar c}^A \zeta^{\cal I} \label{bosetransformazie}\\
\delta {\zeta}_{{\cal I}}&=& {\rm i}\, {\cal U}^{{\cal J} B}_u \, \nabla_\mu q^u \, \gamma^\mu  c^A \, \varepsilon_{AB}
\,
\IC_{{\cal I}{\cal J}}  \label{fermitransformazie1}\\
\delta {\zeta}^{{\cal I}}&=& {\rm i}\, {\cal U}_{{\cal J} B\vert u} \, \nabla_\mu q^u \, \gamma^\mu  c_A \,
\varepsilon^{AB} \, \IC^{{\cal I}{\cal J}} \label{fermitransformazie2}
\end{eqnarray}
where the anticommuting SUSY parameters have being denoted $c_A$ and $c^A$. These are chiral spinors $\gamma_5 c_A \, =
\, c_A$, $\gamma_5 c^B\, = \,-\, c^B$ that transform in the fundamental $\frac{1}{2}$ representation of
$\su(2)_\mathrm{I}$.
\par
The quantum number assignments of all the fields of the generic $\mathcal{N}=2$ sigma model are summarized in table
\ref{topotable} where the Euclidian spin group has been split into its left and right factors:
\begin{equation}\label{splittospin}
    \so(4)_{spin} \, = \, \su(2)_\mathrm{L} \oplus \su(2)_\mathrm{R}
\end{equation}
\begin{table}[h!]
\begin{center}
\begin{tabular}{|c||c||c||c||c||c||c||}
\hline Field   & $\mathrm{SU(2)_L}$  & $\mathrm{SU(2)_R}$  & $\mathrm{SU(2)_I}$ & $\mathrm{USp(2m)}$ & $\mathrm{R}$-sym & degree\\
 name   & rep.  & rep.  & rep. & rep. & $\mathrm{R}$ & d \\
\hline
$q^u$ & $0$ & $0$ & $0$ &$0$ &$0$ & 0   \\
\hline
$\mathcal{U}^{A \mathcal{I}}$ & $0$ & $0$ & $\frac 12$ &$\mathbf{2m}$ & 0 & 1   \\
\hline
$\mathcal{U}_{A \mathcal{I}}$ & $0$ & $0$ & $\frac 12$ &$\overline{\mathbf{2m}}$ & 0 & 1   \\
\hline
$\zeta^\alpha_{\mathcal{I}}$ & $\frac 12$ & $0$ & $0$ &$\overline{\mathbf{2m}}$ & -1 & 0   \\
\hline
$\zeta_{\dot{\alpha}}^{\mathcal{I}}$ & $0$ & $\frac 12$ & $0$ &${\mathbf{2m}}$ & 1 & 0   \\
\hline \hline
\end{tabular}
\caption{\sl Quantum Number assignments in the $\mathcal{N}=2$ sigma model \label{topotable}}
\end{center}
\end{table}
In such a table we display also the charges with respect to so named $\mathrm{R}$-symmetry \cite{Fayet:1976cr} that is the phase symmetry defined by rotating the supersymmetry parameters in the following way:
\begin{equation}\label{ruttosym}
    c_A \, \to \, e^{{\rm i} \theta} \, c_A \quad ; \quad c^A \, \to \, e^{-{\rm i} \theta} \, c^A
\end{equation}
The transformation equations (\ref{bosetransformazie},\ref{fermitransformazie1},\ref{fermitransformazie2}) and the
action (\ref{kinerige}) remain invariant if all the fields are rotated according to:
\begin{equation}\label{ruttosym2}
    \mbox{field} \, \to \, e^{{\rm i}\, R_{field} \theta} \, \mbox{field}
\end{equation}
The next step of this summary consists of the topological twist of the $\mathcal{N}=2$-theory that we have presented.
\par
Before performing the formal manipulations that lead to such a twist it is convenient to analyze a crucial rewriting of
the bosonic action:
\begin{eqnarray}
\mathcal{A}^{\mathcal{N}=2}_{Bose} & = & \int \, h_{uv} \,
\partial_\mu q^u \, \partial_\nu  q^v \, g^{\mu\nu}  d^4x\label{BoseLag}
\end{eqnarray}
which constitutes the real motivation to reinterpret this sigma-model as a topological field theory.
\par
To this effect let us introduce the following $4\times 4m$ bivector:
\begin{equation}\label{triholequa}
    \mathcal{E}_\mu^{\phantom{\mu}u} \, \equiv \,\partial_\mu q^u \, - \, \mathrm{O}_{xy} \, j^{x|\rho}_\mu \,
    \partial_\rho q^s \,  J^{y|u}_s
\end{equation}
where, according to (\ref{afeq4}), $j^x$ and $J^y$ denote the tripletû of
quaternionic complex structures, respectively of the base manifold $\mathcal{M}_4$ and of the target manifold
$\mathcal{N}_{4m}$ and where $\mathrm{O} \in \mathrm{SO(3)}$ is an arbitrary orthogonal matrix in three-dimensions.
Furthermore let us recall the following two identities:
\begin{eqnarray}
  h_{uv} \, J^{x|u}_{s} \,  J^{y|v}_{t}&=& \delta^{xy}\, h_{st} \,- \, \epsilon^{xy\ell} \, K^{\ell}_{st} \nonumber\\
g^{\mu\nu} \, j^{\,x|\rho}_{\mu} \,  j^{\,y|\sigma}_{\nu}&=& \delta^{xy}\, g^{\rho\sigma} \,- \, \epsilon^{xy\ell} \,
k^{\ell|\rho\sigma}\label{identicus}
\end{eqnarray}
where  $g_{\mu\nu}$ denotes the HyperK\"ahler metric of the base manifold $\mathcal{M}_4$, $h_{uv}$ that of the target
manifold $\mathcal{N}_{4m}$ and where $k^{\ell}_{\rho\sigma}$ and $K^{\ell}_{st}$ are the components of the triplets of
HyperK\"ahler 2-forms $\mathbf{k}^\ell$ and $\mathbf{K}^\ell$ respectively defined on the two manifolds
$\mathcal{M}_4$,$\mathcal{N}_{4m}$. Using (\ref{identicus}), by means of a straightforward calculation one can verify
that if we set:
\begin{equation}\label{perfectquadro}
\parallel \mathcal{E}_\mu^{\phantom{\mu}u} \parallel^2 \, \equiv \, h_{uv} \, g^{\mu\nu}
\,{\mathcal{E}}_\mu^{\phantom{\mu}u} \,{\mathcal{E}}_\nu^{\phantom{\mu}v}
\end{equation}
we obtain:
\begin{eqnarray}\label{confiteorDei}
    \parallel \mathcal{E}_\mu^{\phantom{\mu}u} \parallel^2 & = &
    \left(1\, +\, \mbox{Tr} \left[\mathrm{O}^T \,\mathrm{O} \right]\right) \,{\cal L}^{\mathcal{N}=2}_{Bose}
\, + \, 4 \, \mathrm{O}_{xy}  \,\left( \mathbf{k}^x\right)^{\mu\nu} \,\left(  q^\star \mathbf{K}^y\right)_{\mu\nu}\nonumber\\
\left(  q^\star \mathbf{K}^y\right)_{\mu\nu} & \equiv & K^{y}_{uv} \, \partial_\mu q^u \partial_\nu q^v
\end{eqnarray}
where $ q^\star \mathbf{K}^x$ denotes the pull-back of the HyperK\"ahler forms of the target manifold onto
the base-manifold by means of the map $ q$. We further take into account that for any pair of $2$-forms
$\omega^{[2]}$ and $\pi^{[2]}$ we have:
\begin{equation}\label{oppi}
    \int \, \omega^{\mu\nu} \,\pi_{\mu\nu}\,\times \,\mathrm{Vol_4}  \, = \, \int \, \star_g \, \omega^{[2]}\wedge\pi^{[2]}
\end{equation}
where $\star_g$ denotes the Hodge-dual with respect to the metric $g$ and that the HyperK\"ahler forms of the base
manifold are chosen antiselfdual:
\begin{equation}\label{antiselfi}
    \star_g \,\mathbf{ k}^x \, = \, - \, \mathbf{ k}^x
\end{equation}
Then from eq.(\ref{confiteorDei}) we conclude:
\begin{equation}\label{quadratabile}
    \mathcal{A}^{\mathcal{N}=2}_{Bose} \, = \, \frac{1}{8} \,\underbrace{\mathrm{O}_{xy} \,
    \int_{\mathcal{M}_4} \, \mathbf{k}^x \wedge  q_\star \mathbf{K}^y}_{\mathcal{A}_{top}[q,\mathrm{O}]}
    \,+ \, \frac{1}{4} \underbrace{\int_{\mathcal{M}_4} \parallel \mathcal{E}_\mu^{\phantom{\mu}u} \parallel^2 \,\times
    \mathrm{Vol_4}}_{\mathcal{A}_{\mathrm{Hyp}}[q,\mathrm{O}]}
\end{equation}
Eq.(\ref{quadratabile}) tells us that the classical action of the purely bosonic
sigma-model is the sum of two terms. The first term is  topological, independent from the base-manifold metric and from
the continuous deformation of the map $q$ within the same \textit{homotopy class}. The second term is the integral of a
perfect square. It follows that within each \textit{homotopy class}, the classical action has an absolute minimal
attained by those configurations that correspond to $\mathcal{E}_\mu^{\phantom{\mu}u}\, = \,0$. Looking back at eq.
(\ref{triholequa}) we see that the vanishing of such a structure is precisely the condition of
triholomorphicity discussed in the introduction (see eq. (\ref{afeq4})) which  defines the
\textit{hyperinstantons}.
\par
There are still two relevant mathematical questions to be clarified:
\begin{description}
  \item[a)] How and to what degree of definiteness the topological number:
  \begin{equation}\label{paronzo}
    \mathcal{A}_{top}[q,\mathrm{O}]\, = \,\mathrm{O}_{xy} \,
    {\int_{\mathcal{M}_4} \, \mathbf{k}^x \wedge  q_\star \mathbf{K}^y}
  \end{equation}
classifies the homotopy classes of the maps: $ q \, : \, \mathcal{M}_{4} \, \to \,\mathcal{N}_{4m}$?
  \item[b] What are the true indipendent choices of the orthogonal matrix $\mathrm{O}$ up to diffeomorphisms or other
  symmetries?
\end{description}
Postponing the important discussion of the above two points to later sections, we just observe that the topological
interpretation of the sigma-model is effective and goes along classical lines if we succeed in proving that:
\begin{enumerate}
  \item There is a suitable topological BRST-charge $s$ with respect to which the topological term is BRST-closed but
  not exact:
  \begin{equation}\label{closure}
    s \,\mathcal{A}_{top}[q,\mathrm{O}]\, = \,0 \quad ; \quad (\mbox{BRST-cocycle})
  \end{equation}
  \item There exists a suitable gauge fermion $\Psi_{gauge}$, such that
  \begin{equation}\label{fuceccus}
    \mathcal{A}_{\mathrm{Hyp}}[q,\mathrm{O}]\, + \, \mbox{ghost - antighost terms} \, = \, s \,\Psi_{gauge}
  \end{equation}
\end{enumerate}
Under the above conditions we can interpret $\mathcal{A}_{top}[q,\mathrm{O}]$ as the classical action of a topological field
theory and $\mathcal{A}_{\mathrm{Hyp}}[q,\mathrm{O}]$ as the leading terms of a BRST gauge fixing of the topological symmetry.
With the usual argument that being BRST-exact the term $\mathcal{A}_{\mathrm{Hyp}}[q,\mathrm{O}]$ can be multiplied by an
arbitrary constant $t\to \infty$ we obtain localization of the functional integral on the hyperinstanton configurations. In
particular all this is assured if the hyperinstanton equation $\mathcal{E}_\mu^{\phantom{\mu}u}$ happens to be the BRST-variation
of a suitable antighost.
\par
The above outlined programme is realized by the topological twist of the 4D $\mathcal{N}=2$ sigma-model: we  describe it
in detail below. Before doing that we want to emphasize a very crucial detail that will turn out to be at the core of all
our results.
\par
Let us first consider the standard mechanism behind the BRST-invariance of a topological action such as
(\ref{paronzo}). Usually we have that the integrand, in our case $\mathfrak{I}^{(4,0)}\, = \, \mathrm{O}_{xy}
\,\,\mathbf{k}^x \wedge  q_\star \mathbf{K}^y$, is a top-form satisfying a descent equation of the form:
\begin{equation}\label{discesona}
    \mathrm{s} \,\mathfrak{I}^{(4,0)} \, = \,\mathrm{d} \,\mathfrak{I}^{(3,1)}
\end{equation}
where the labels $(d,g)$ specify the degree as a differential form and the ghost number of the labeled object.
Eq.(\ref{discesona}) is true also in the case of the hyperinstantons we consider in this paper. Then we obtain:
\begin{equation}\label{farmalucco}
    \mathrm{s} \,\int_{\mathcal{M}_4} \, \mathfrak{I}^{(4,0)} \, = \, \int_{\mathcal{M}_4} \,
    \mathrm{d} \,\mathfrak{I}^{(3,1)} \, = \,\int_{\partial \mathcal{M}_4} \,\mathfrak{I}^{(3,1)}
\end{equation}
If the base manifold has no boundary $\partial \mathcal{M}_4\, = \, 0$, the last integral in eq.(\ref{farmalucco})
vanishes and the topological action is BRST invariant. However it is an intrinsic crucial feature of our hyperintantons
 that the base manifold is just $\mathbb{R}_+\times \mathrm{T^3}$, which has a
boundary. Hence the last integral in eq.(\ref{farmalucco}) has to be considered. If the boundary conditions on the fields are
such that it vanishes, then the topological action is BRST invariant and the topological field theory obtained by the twist
procedure stands on its feet. The condition for BRST invariance is thus clearly formulated.
\subsection{Detailed derivation of the topological twist}
 We assume that the target HyperK\"ahler manifold has the reduced holonomy mentioned in eq.(\ref{olonomiaRest}). Then all the
fields can be classified with the following quantum numbers:
\begin{equation}\label{collassus}
    \null^R\left(\mathrm{L,R,I,Q}\right)_d
\end{equation}
where $R$ denotes the $R$-charge, $d$ denotes the form degree and $\mathrm{L,R,I,Q}$ denote the representations with
respect to the $\mathrm{SU(2)}$-groups, $\mathrm{L,R,I,Q}$.
\begin{table}[h!]
\begin{center}
\begin{tabular}{|c||c||c||c||c||c||c||c||}
\hline Field   & $\mathrm{SU(2)_L}$  & $\mathrm{SU(2)_R}$  & $\mathrm{SU(2)_I}$& $\mathrm{SU(2)_Q}$  &
$\mathrm{SO(m)}$ & $\mathrm{R}$-sym & degree \\
 name   & rep.  & rep.  & rep. & rep. & rep. & $\mathrm{R}$ & d \\
\hline
$q^u$ & $0$ & $0$ & $0$ &$0$ &$0$ & $0$ &$0$  \\
\hline
$\mathcal{U}^{\phantom{k}A}_{k\phantom{A}\dot{B}}$ & $0$ & $0$ & $\frac 12$ & $\frac 12$& $\mathbf{m}$ & 0 & 1   \\
\hline
$\mathcal{U}^{\phantom{kA}\dot{B}}_{k A}$ & $0$ & $0$ & $\frac 12$ & $\frac 12$& $\mathbf{m}$ & 0 & 1   \\
\hline
$\zeta^{\alpha \dot{A}}_{k}$ & $\frac 12$ & $0$ & $0$ & $\frac 12$& $\mathbf{m}$ & -1 & 0   \\
\hline
$\zeta_{\dot{\alpha}\dot{A}| k}$ & $0$ & $\frac 12$ & $0$ & $\frac 12$& $\mathbf{m}$ & 1 & 0   \\
\hline \hline
\end{tabular}
\caption{\sl Quantum Number assignments in the $\mathcal{N}=2$ sigma model with restricted holonomy. The indices
$\alpha,\beta\, =\, 1,2$ span the fundamental representation of $\su(2)_\mathrm{L}$. The indices
$\dot{\alpha},\dot{\beta} \, = \, \dot{1},\dot{2}$, span the fundamental representation of $\su(2)_\mathrm{R}$. The
indices $A,B \, = \, 1,2$, span the fundamental representation of $\su(2)_\mathrm{I}$.The indices $\dot{A},\dot{B} \, =
\, \dot{1},\dot{2}$, span the fundamental representation of $\su(2)_\mathrm{Q}$.\label{balanda}}
\end{center}
\end{table}
\par
The explicit charge assignments are mentioned in table \ref{balanda}.
\par
According to \cite{Anselmi:1993wm} the topological twist is performed by means of the following  formal steps:
\begin{enumerate}
  \item The spin of the fields is redefined by means of an identification of the original spin group with the bundle
  structural group $\so(4)_{\mathrm{IQ}}$ defined in eq.(\ref{meconsuelo}). Explicitly one sets:
  \begin{eqnarray}\label{caraturus}
    \so(4)_{spin}^\prime & = & \su(2)_\mathrm{L}^\prime \, \oplus \, \su(2)_\mathrm{R}^\prime \nonumber\\
 \su(2)_\mathrm{L}^\prime & = & \mbox{diag} \left[\su(2)_\mathrm{L} \, \oplus \, \su(2)_\mathrm{Q}\right]\nonumber\\
 \su(2)_\mathrm{R}^\prime & = & \mbox{diag} \left[\su(2)_\mathrm{R} \, \oplus \, \su(2)_\mathrm{I}\right]
  \end{eqnarray}
 \item One identifies the $\mathrm{R}$-symmetry charge with the ghost number $g$.
 \item According to the above scheme, after the twist the quantum numbers that characterize any field are
 $\left(\mathrm{L^\prime,R^\prime}\right)^g_d$ and we have:
 \begin{equation}\label{fanciullaWest}
    \null^R\left(\mathrm{L,R,I,Q}\right)_d \, \rightarrow \, \left(\mathrm{L+Q},\mathrm{R+I}\right)^R_d \,
    \equiv \,\left(\mathrm{L}^\prime,\mathrm{R}^\prime\right)^g_d
 \end{equation}
\item Decomposing the supersymmetry parameters into irriducible representations before and after the twist
 we have:
\begin{eqnarray}
  c^{A\dot{\beta}} &=& \left(0, \ft 12,\ft 12,0\right) \, \to \, \left(0, \ft 12+\ft 12\right)\, = \,
  \underbrace{\left(0, 0\right)}_{BRST param.}\, + \, \left(0, 1\right)  \\
  c_{A\alpha} &=& \left(\ft 12, 0,\ft 12,0\right) \, \to \, \left(\ft 12, \ft 12\right)
\end{eqnarray}
and we identify the BRST-charge as the operator corresponding to the unique scalar supersymmetry parameter with
respect to the redefined spin group. In practice the BRST algebra is obtained from the supersymmetry transformation
rules (\ref{bosetransformazie},\ref{fermitransformazie1}, \ref{fermitransformazie2}) by setting:
\begin{equation}\label{practicus}
    c^{A\dot{\beta}} \, = \, \epsilon^{A\dot{\beta}} \, \underbrace{\mathbf{e}}_{\begin{array}{c}
                                                                                   BRST \\
                                                                                   param.
                                                                                 \end{array}}\quad ; \quad
     c_{A{\beta}}\, = \, 0
\end{equation}
\end{enumerate}
\begin{table}[h!]
\begin{center}
\begin{tabular}{|c||c||c||c||c||c||c||}
\hline Field   & $\mathrm{SU(2)_L}^\prime$  & $\mathrm{SU(2)_R}^\prime$  &
$\mathrm{SO(m)}$ & ghost number & degree & interpret.\\
 name   & rep.  & rep.  & rep. &  g & d & \null\\
\hline
$q^u$ & $0$ & $0$ & $0$ &$0$ &$0$ & phys. field  \\
\hline
$\mathcal{U}^{\phantom{k}A}_{k\phantom{A}\dot{B}}$ & $\ft 12$ & $\ft 12$ & $\mathbf{m}$ & 0 & 1 & phys.field   \\
\hline
$\mathcal{U}^{\phantom{kA}\dot{B}}_{k A}$ &  $\frac 12$ & $\frac 12$& $\mathbf{m}$ & 0 & 1 & phys. field  \\
\hline
$\zeta^{\alpha \dot{A}}_{k}$ & $\frac 12+ \ft 12$ & $0$ & $\mathbf{m}$ & -1 & 0 & antighost  \\
\hline
$\zeta_{\dot{\alpha}\phantom{A}| k}^{\phantom{\alpha}\dot{A}}$ &$\frac 12$ &  $\frac 12$& $\mathbf{m}$ & 1 & 0 & top ghost  \\
\hline \hline
\end{tabular}
\caption{\label{baltaruca}\sl Quantum Number assignments of all the fields after topological twist.}
\end{center}
\end{table}
The quantum numbers of all the fields after the twist and their interpretation within the BRST complex are displayed in
table \ref{baltaruca} and their BRST transformations take the following explicit form:
\begin{eqnarray}
  \mathcal{U}^{\phantom{k}A}_{u|k\phantom{A}\dot{B}} \, s q^u  &=& \epsilon^{A\dot{\beta}} \, \zeta_{\dot{\beta} \dot{B}|k}
  \label{deformio} \\
  s \, \zeta_{\dot{\beta} \dot{B}|k} &=& 0 \label{buonfantasma}\\
  s \, \zeta_{k}^{\alpha \dot{A}}&=& {\rm i} \left( e^\dagger_\mu \,
  e_\nu\right)^\alpha_{\phantom{\alpha}\dot{C}}\epsilon^{\dot{A}\dot{C}} E^\nu_{k|u} \partial^\mu q^u
  \label{antiagosto}
\end{eqnarray}
In the last of the above equations we made use of the explicit form of the gamma matrix basis introduced in
eq.(\ref{gammatine}).
\par Next raising the index $\dot{A}$ as provided by the epsilon-symbol, we conclude that the variation of the
antighost has the following appearance:
\begin{equation}
    s \, \zeta_{k}^{\alpha \dot{A}}\,= \, {\rm i} \left( e^\dagger_\mu \,
  e_\nu\right)^{\alpha\dot{A}} E^\nu_{k|u} \partial^\mu q^u
  \label{antiluglio}
\end{equation}
There are two cases in the decomposition of a two-index tensor $t^{\alpha \dot{A}}$, the antisymmetric case that counts
one degrees of freedom and the symmetric case that counts three degrees of freedom. Hence we can set:
\begin{equation}\label{falascus}
    \zeta_{k}^{\alpha\dot{A}}\, = \, \epsilon^{\alpha\dot{A}} \,\zeta^\bullet_{k} \, + \,{\rm i} \,
    \sigma_x^{\alpha\dot{A}} \, \zeta^x_{k}
\end{equation}
Projecting onto the two cases we get:
\begin{eqnarray}
  s\,\zeta^\bullet_{k}  &\propto & g_{\mu\nu}\, E^\nu_{k|u} \partial^\mu q^u \label{coriandolo1}\\
   s\,\zeta^x_{k} &\propto & J^{+x}_{\mu\nu} \, E^\nu_{k|u} \partial^\mu q^u \label{coriandolo2}
\end{eqnarray}
If we introduce the matrix:
\begin{equation}\label{mentuccia}
    A^{\nu\mu}_k \, = \, E^\nu_{k|u} \partial^\mu q^u
\end{equation}
vanishing of the BRST variation of the two antighosts $\zeta^\bullet_{k}$ and $\zeta^x_{k}$, which is what defines the
topological gauge fixing, implies the two conditions:
\begin{eqnarray}\label{borlotto}
    A_{k}^{\nu\mu} g_{\mu\nu} & = & 0 \label{borlotto}\\
     A_{k}^{[\nu\mu]} \, + \, \ft 12 \,\epsilon^{\mu\mu\rho\sigma} \,
    A_{k}^{[\rho\sigma]}& =& 0 \label{borlottoDue}
\end{eqnarray}
Indeed the self-dual character  of $J^{+x}_{\mu\nu}$ in eq.(\ref{coriandolo2}) yields the result that the self-dual
part of the tensor $A_{k}^{[\nu\mu]}$ is set to zero.
\par
In \cite{Anselmi:1993wm} it was shown that eq.s (\ref{borlotto}) and (\ref{borlottoDue}) are equivalent to the
statement:
\begin{equation}\label{circusfissu}
    A_{k}\, - \, J^{x} \circ A_k \circ J^x \, = \,0
\end{equation}
which on its turn is the same as the triholomorphicity condition:
\begin{equation}\label{lomonosov}
    \partial^\mu q^u \, - \,  \, J^{-x|\mu\rho} \,
    \partial_\rho q_s \,  J^{-x|su} \, = \,0
\end{equation}
In the case of flat HyperK\"ahler manifold eq.(\ref{lomonosov}) is alternatively rewritten as (\ref{borlottoDue}) or:
\begin{equation}
\matrix{\partial_\mu q_\mu=0,&
\partial_\mu q_\nu-\partial_\nu q_\mu+\varepsilon_{\mu\nu\rho\sigma}
\partial^\rho q^\sigma=0.}
\label{flat}
\end{equation}
Here we note that the second equation can be viewed as a self-duality condition of the field strength $
F_{\mu\nu} = \partial_\mu q_\nu - \partial_\nu q_\mu$ and the first equations resemble a gauge fixing for the
potential $q_\mu$. We would like to underlying that the latter is not a choice since that equation stems from the triholomorphic map condition and it can not be changed. This point is clearly discussed also in \cite{Anselmi:1993wm}.

\section{Triholomorphic hyperinstantons and Beltrami vector fields}
Having clarified the local and global structure of both the base and the target space, we conclude that the maps
\begin{equation}\label{hypinst}
    q \,:\, \mathcal{M}_4 \,\to \,\mathcal{N}_4
\end{equation}
 that constitute the functional space of our considered sigma-model have to be periodic up to the lattice $\Lambda$,
namely we must have:
\begin{equation}
\forall \mathbf{v} \, \in \, \Lambda \quad : \quad\left\{ \begin{array}{rcl} q^0(\mathrm{U},\mathbf{X}+\mathbf{v}) &=&
q^0(\mathrm{U},\mathbf{X}) \,\quad ; \quad |q^0(\mathrm{U},\mathbf{X})| < \infty \\
\mathbf{q}(\mathrm{U},\mathbf{X}+\mathbf{v}) &=& \mathbf{q}(\mathrm{U},\mathbf{X})\,\quad ; \quad
|\mathbf{q}(\mathrm{U},\mathbf{X})| < \infty
\end{array}\right.
\end{equation}
where we have denoted $\mathbf{X}=\{\mathrm{X,Y,Z}\}$ and $\mathbf{q}=\{\mathrm{q^1,q^2,q^3}\}$. We will discuss below
how we can implement the above boundary condition in our functional space.
\par
The main point of this section is to show that the hyperinstanton equation (\ref{afeq4}) with $\mathrm{O}\,= \,
\mathbf{1}$ reduces to Beltrami equation (\ref{formaduale}) for vector fields in $D=3$.
\par

We assume the following ansatz:
\begin{eqnarray}
\label{facto}
\frac{\mathbf{q}(U, \mathbf{X})}{q_0(U, \mathbf{X})} = \mathbf{T}(\mathbf{X})\,,
\end{eqnarray}
namely the ratio between the ``spatial" components $\mathbf{q}$ and the ``time" component is independent of the time coordinate $U$
and it is a periodic function of $\mathrm{T^3}$ (the coordinates $\mathbf{q}/q_0$ are coordinates on a projective space, we can
view the space as a cone over $\mathrm{T^3}$). We can solve this ansatz by setting
\begin{eqnarray}
\label{factoA} q_0(\mathrm{U}, \mathbf{X}) = f(\mathrm{U}) \, G(\mathbf{X})\,, ~~~~~~~ \mathbf{q}(\mathrm{U}, \mathbf{X}) =
f(\mathrm{U}) \, \mathbf{H}(\mathbf{X})\,, ~~~~~
 \end{eqnarray}
Equations (\ref{lomonosov}) become
\begin{eqnarray}
\label{factoB}
\partial_\mathrm{U} q_0 + {\mathbf\nabla} \cdot  \mathbf{q}=0\,, ~~~~~~~
\partial_\mathrm{U} \mathbf{q} - \mathbf{\nabla} q_0 + \mathbf{\nabla} \times \mathbf{q}=0\,.
\end{eqnarray}
The gradient $\nabla$ is taken over the spatial coordinates, namely on the base manifold.
Inserting the ansatz (\ref{factoA}) into (\ref{factoB})
we get the following equations
\begin{eqnarray}
\label{factoC} f^{-1} \partial_\mathrm{U} f  = - G^{-1} \nabla \cdot \mathbf{H} =0\,, ~~~~~~~ (f^{-1} \partial_\mathrm{U} f)
\mathbf{H} - \nabla G +   \mathbf{\nabla} \times \mathbf{H} =0\,.
\end{eqnarray}
In the first equation, by separation of variables, we set $f^{-1} \partial_U f  = - \mu$ with $\mu >0$, leading to $f(\mathrm{U})
= K e^{- \mu \mathrm{U}}$ defined on $\mathbb{R}_+$. Thus, the two equations read
\begin{eqnarray}
\label{factoD}
\nabla \cdot \mathbf{H} = \mu \, G\,, ~~~~~~~
\mathbf{\nabla} \times \mathbf{H}  - \mu\, \mathbf{H} = \nabla G\,,
\end{eqnarray}
Acting with $\nabla$ on the second equation we get $- \mu \nabla \cdot \mathbf{H} = \nabla^2 G$ which implies
\begin{eqnarray}
\label{factoE}
\nabla^2 G = - \mu^2 G\,.
\end{eqnarray}
In addition, we can redefine $\mathbf{H}$ as
\begin{eqnarray}
\label{factoF}
\mathbf{H} = \mathbf{Y} - \frac{1}{\mu} \nabla G\,,
\end{eqnarray}
to get
\begin{eqnarray}
\label{factoG}
\nabla \cdot \mathbf{Y} = 0\,, ~~~~~~~
\nabla \times \mathbf{Y} - \mu \mathbf{Y} =0\,.
\end{eqnarray}
The first equation is a consequence of the second equation (assuming that $\mu \neq 0$) and the latter
is the vectorial version of the Beltrami equation for the $1$-form $\mathbf{Y}_{[1]}$ discussed in the
introduction. Written in form language, (\ref{factoG}) are
\begin{eqnarray}
\label{factoH} \mathrm{d} \star \mathbf{Y}_{[1]} = 0\,, ~~~~~~~~~ \star \mathrm{d} \mathbf{Y}_{[1]} - \mu \mathbf{Y}_{[1]} =0\,.
\end{eqnarray}
In this way, we have proven that all ``conical'' solutions with the ansatz (\ref{facto}) are in correspondence with solutions of
the Beltrami differential equation (\ref{factoH}).
\par
If $\partial_\mathrm{U} f =0$, namely if $\mu =0$, we get a different solution
\begin{eqnarray}
\label{factoI}
\mathbf{H} = \nabla \times \mathbf{A}\,, ~~~~~~
G = \nabla \cdot \mathbf{A}\,, ~~~~~
\nabla^2 \mathbf{A} =0\,.
\end{eqnarray}
where the last equation is the Laplace equation on the torus $\mathrm{T^3}$, whose solutions are very well-known.
\par
In conclusion, in view of the above discussion we can write the general form of the triholomorphic hyperinstantons:
\begin{equation}\label{copinus}
    q \, : \, \mathbb{R}_+ \times \mathrm{T^3} \, \rightarrow \, \mathbb{R}^4
\end{equation}
in the following form:
\begin{equation}\label{tonnocarciofinus}
    q(\mathrm{U},\mathbf{X}) \, = \, \sum_{\mu \in \mathfrak{S}} \, \sum_{\mathfrak{m} \in \mathfrak{M}(\mu)} \,
    q_{[\mu]}\left(\mathrm{U},\mathbf{X}|\mathfrak{m}\right)
\end{equation}
where $\mathfrak{S}$ denotes the spectrum of eigenvalues $\mu$ of the $\star_g \, \mathrm{d}$ operator on $\mathrm{T^3}$, whose
squares $\mu^2$ are eigenvalues of minus the Laplacian on the same space, and $\mathfrak{M}(\mu)$ denotes the parameter space of
solutions of eq.s (\ref{factoH}) and (\ref{factoF}) combined together.
 Setting:
\begin{equation}
\begin{array}{rclcrcl}
  \Box_{\mathrm{T^3}}\mathfrak{W}_{[\mu]}^{\ell}\left(\mathbf{X}\right)&=&-4\, \mu^2 \, \mathfrak{W}_{[\mu]}^{\ell}
  \left(\mathbf{X}\right) & ; & \ell & = & 1,\dots , \delta_\mu \\
 \star \, \mathrm{d}_{\mathrm{T^3}} \mathbf{Y}^{[\mu|I]}\left(\mathbf{X}\right) &=& \mu \,\mathbf{Y}^{[\mu|I]}\left(\mathbf{X}\right)& ; &
 I & = & 1,\dots , d_\mu \\
\end{array}
\end{equation}
where $\delta_\mu$ and $d_\mu$ are the degeneracies, respectively of the Laplacian eigenvalue and of the $\star
\mathrm{d}$ eigenvalue, we can write the general solution as follows:
\begin{eqnarray}
  \Phi_{[\mu]}\left(\mathrm{U},\mathbf{X}\right)&=& e^{ -2\,\mu \,\mathrm{U}}\, \sum_{\ell = 1}^{\delta_\mu}\mathfrak{b}_{\mu,\ell} \,
  \mathfrak{W}_{[\mu]}^{\ell}\left(\mathbf{X}\right) \nonumber\\
  \Omega_{[\mu]}\left(\mathrm{U},\mathbf{X}\right)&=&e^{-\,2\,\mu \,\mathrm{U}}\,
  \sum_{I = 1}^{d_\mu}\mathfrak{c}_{\mu,I} \,
  \mathbf{Y}^{[\mu|I]}\left(\mathbf{X}\right) \nonumber\\
  {q^0_{[\mu]}}\left(\mathrm{U},\mathbf{X}|\mathfrak{m}\right)&=& \, - \,\partial_\mathrm{U}
  \,\Phi_{[\mu]}\left(\mathrm{U},\mathbf{X}\right)\nonumber\\
{\mathbf{q}_{[\mu]}}\left(\mathrm{U},\mathbf{X}|\mathfrak{m}\right)&=&
  \,\nabla \,\Phi_{[\mu]}\left(\mathrm{U},\mathbf{X}\right)\, +\, \Omega_{[\mu]}\left(\mathrm{U},\mathbf{X}\right)\label{generasolata}
\end{eqnarray}
where  $\mathfrak{c}_{\mu,I}$ are constant parameters. In comparison with equations
(\ref{factoA},\ref{factoB},\ref{factoC},\ref{factoD}) we see that we can identify $G(\mathbf{X})\, = \,\sum_{\ell =
1}^{\delta_\mu}\mathfrak{b}_{\mu,\ell} \,
  \mathfrak{W}_{[\mu]}^{\ell}\left(\mathbf{X}\right)$, $f(U) \, = \, e^{- 2\, \mu}$, $\mathbf{Y}(\mathbf{X})\, = \,\sum_{I = 1}^{d_\mu}
  \mathfrak{c}_{\mu,I} \,
  \mathbf{Y}^{[\mu|I]}\left(\mathbf{X}\right)$. The parameters $\mathfrak{m}$ are given by the union of the $\mathfrak{c}_{\mu,I}$ with
  the $\mathfrak{b}_{\mu,\ell}$.
\par
The spectrum $\mathfrak{S}$ of the operator $\star \mathrm{d}_{\mathrm{T^3}}$ was completely calculated in \cite{Fre:2015mla} and
organized into irreducible representations of a Universal Classifying Group. An easy by-product of that calculation and of that
classification is the derivation of the spectrum of the laplacian operator. In fact we just find that the degeneracies of the two
operators are the same at equal value of $\mu$:
\begin{equation}\label{carlovaggio}
    d_\mu \, = \, \delta_\mu
\end{equation}
Hence the total dimension of the parameter space  $\mathfrak{M}(\mu)$ of a $\mu$-hyperinstanton is $2\times d_\mu$.
\subsection{The standard Fourier expansion on $\mathrm{T^3}$ and the condition of triholomorphicity}
\label{comparazia}
 In the previous subsection we have seen that, under the mild condition (\ref{factoA}), the first order differential constraint of
triholomorphicity (\ref{lomonosov}) is equivalent  to Beltrami equation (\ref{formaduale}) on the three-torus; in this way the
enumeration of hyperinstantons can be reduced to the doubled enumeration of solutions of the first order differential equation of
Beltrami. This provides a powerful tool to organize the functional integration localized on the hyperinstantons into a discrete
sum over the spectrum $\mathfrak{S}$ of the $\star_g \, \mathrm{d}$ operator plus a finite dimensional integration over the
parameter space of hyperinstantons at fixed eigenvalue $\mu$. Such an organization of the functional integration will be
illustrated in the next subsection \ref{organisazia}. In the present one, in order to appreciate the field theoretical meaning of
that construction we consider the standard expansion of the scalar fields $q^u(\mathrm{U},\mathbf{X})$ into Fourier modes on the
three-torus and we analyse the triholomorphic constraint in momentum space.
\par
Interpreting the coordinate $\mathrm{U}$ as the euclidian time after Wick rotation, the $4$ coordinates $q^u$ of the target space
are just $4$ scalar fields quantized with periodic boundary condition in a cubic box (namely $\mathrm{T^3}$) and in full
generality we can write the discrete Fourier expansion:
\begin{equation}\label{fourietto1}
    q^u(\mathrm{U},\mathbf{X})\, = \, \sum_{\mathbf{k}\in \Lambda} \, \exp\left[ 2 \,\pi\, {\rm i}\, \mathbf{k}\cdot
    \mathbf{X}\right] \, a^u_{\mathbf{k}}(\mathrm{U})
\end{equation}
where $\Lambda$ is the cubic lattice given by all momentum three-vectors $\mathbf{k}\, = \, \{k_x,k_y,k_z\}$ whose components are
integer valued: $k_{x,y,z} \in \mathbb{Z}$. The reality of the scalar fields $q^u$ yields the standard condition:
\begin{equation}\label{colazione1}
    a^u_{-\mathbf{k}}(\mathrm{U})\, = \, a^u_{\mathbf{k}}(\mathrm{U})^\star
\end{equation}
Hence to each momentum pair $(\pm \mathbf{k})$ we associate $4\times 2 \, = \, 8$ real parameters depending on the \textit{time}
$\mathrm{U}$:
\begin{equation}\label{colazione2}
    a^u_{\mathbf{k}}(\mathrm{U})\, = \, \alpha^u_{\mathbf{k}}(\mathrm{U})\, + \, {\rm i} \, \beta^u_{\mathbf{k}}(\mathrm{U})
\end{equation}
Let us now impose the classical field equation of the free field:
\begin{equation}\label{classaequa}
    \Box_{\mathbb{R}_+\times \mathrm{T^3}} \,q^u \, = \,\sum_{\mathbf{k}\in \Lambda} \,
    \exp\left[ 2 \,\pi\, {\rm i}\, \mathbf{k}\cdot
    \mathbf{X}\right] \, \left[ -4 \, \pi^2\, \mathbf{k}^2 \, \alpha^u_{\mathbf{k}}(\mathrm{U})\,+
    \, \ddot{\alpha}^u_{\mathbf{k}}(\mathrm{U})\right ] \, = \, 0
\end{equation}
which is solved by setting:
\begin{eqnarray}
  \alpha^u_{\mathbf{k}}(\mathrm{U}) &=& \exp\left[-2\,\pi\, \mu\, \mathrm{U}\right]\, \left(\alpha^{u|+}_{\mathbf{k}}\,+ \, {\rm i}
  \beta^{u|+}_{\mathbf{k}} \right)\, + \,\exp\left[2\,\pi\, \mu\, \mathrm{U}\right]\, \left(\alpha^{u|-}_{\mathbf{k}}\,+ \, {\rm i}
  \beta^{u|-}_{\mathbf{k}} \right) \nonumber\\
  \alpha^u_{-\mathbf{k}}(\mathrm{U}) &=& \exp\left[-2\,\pi\, \mu\, \mathrm{U}\right]\, \left(\alpha^{u|+}_{\mathbf{k}}\,- \, {\rm i}
  \beta^{u|+}_{\mathbf{k}} \right)\, + \,\exp\left[2\,\pi\, \mu\, \mathrm{U}\right]\, \left(\alpha^{u|-}_{\mathbf{k}}\,- \, {\rm i}
  \beta^{u|-}_{\mathbf{k}} \right)\nonumber\\
\end{eqnarray}
where:
\begin{equation}\label{mudefilo}
    \mu \, \equiv \, \sqrt{\mathbf{k}^2}
\end{equation}
Hence for each pair of momentum vectors $\pm \mathbf{k} \in \Lambda$ the general solution of the classical second order field
equation contains $4 \times 4 \, = \, 16$ real parameters given by $\alpha^{u|\pm}_{\mathbf{k}}$ and
$\beta^{u|\pm}_{\mathbf{k}}$.
\par
Let us next consider the constraint of triholomorphicity in Fourier space. Any of the above described classical solutions can be
rewritten as:
\begin{equation}\label{riscrittoFurio}
    q^u \,= \, \sum_{p} \, \exp\left[2\,\pi \, p^\nu \, x_\nu \right] \, a^u (p)
\end{equation}
where:
\begin{eqnarray}\label{momentini}
    p^\nu \, = \, \left\{\omega ,{\rm i}k_x,{\rm i}k_y,{\rm i}k_z\right\} \quad \mbox{with} \quad \omega \, = \, \pm \, \sqrt{\mathbf{k}^2}
\end{eqnarray}
Inserting (\ref{riscrittoFurio}) into the constraint of triholomorphicity (\ref{lomonosov}) we obtain the algebraic condition:
\begin{equation}\label{arkangelsk}
    \mathfrak{E}_{\mu\nu|uv} \, p^\nu \, a^v(p) \, = \, 0
\end{equation}
where the constant numerical tensor $\mathfrak{E}_{\mu\nu|uv} $ is defined as follows:
\begin{equation}\label{ruminante}
    \mathfrak{E}_{\mu\nu|uv}\, \equiv \, \delta_{\mu\nu} \, \delta_{uv} \, + \, \sum_{x=1}^3 J^{-|x}_{\mu\nu} \,J^{-|x}_{uv}
\end{equation}
It follows that for each momentum vector $p^\nu$ the coefficient vector $\vec{a} = a^u(p)$ has to be annihilated by four
matrices:
\begin{equation}\label{coccolino}
    \left(\mathbf{E}_\mu\right)_{uv} \, \equiv \, \mathfrak{E}_{\mu\nu|uv} \, p^\nu \quad ; \quad \mathbf{E}_\mu \cdot \vec{a} \,
    = \, 0
\end{equation}
Let us consider the explicit form of these matrices:
\begin{equation}\label{quattromoschetti}
    \begin{array}{ccccccc}
       \mathbf{E}_0 & = & \left(
\begin{array}{cccc}
 -\omega  &{\rm i} k_x &{\rm i} k_y &{\rm i} k_z \\
-{\rm i} k_x & -\omega  &-{\rm i} k_z &{\rm i} k_y \\
-{\rm i} k_y &{\rm i} k_z & -\omega  &-{\rm i} k_x \\
-{\rm i} k_z &-{\rm i} k_y &{\rm i} k_x & -\omega  \\
\end{array}
\right) & ; & \mathbf{E}_x & = & \left(
\begin{array}{cccc}
{\rm i} k_x & \omega  &{\rm i} k_z &-{\rm i} k_y \\
 -\omega  &{\rm i} k_x &{\rm i} k_y &{\rm i} k_z \\
-{\rm i} k_z &-{\rm i} k_y &{\rm i} k_x & -\omega  \\
{\rm i} k_y &-{\rm i} k_z & \omega  &{\rm i} k_x \\
\end{array}
\right) \\
       \mathbf{E}_y  & = & \left(
\begin{array}{cccc}
{\rm i} k_y &-{\rm i} k_z & \omega  &{\rm i} k_x \\
{\rm i} k_z &{\rm i} k_y &-{\rm i} k_x & \omega  \\
 -\omega  &{\rm i} k_x &{\rm i} k_y &{\rm i} k_z \\
-{\rm i} k_x & -\omega  &-{\rm i} k_z &{\rm i} k_y \\
\end{array}
\right) & ; & \mathbf{E}_z & = & \left(
\begin{array}{cccc}
{\rm i} k_z &{\rm i} k_y &-{\rm i} k_x & \omega  \\
-{\rm i} k_y &{\rm i} k_z & -\omega  &-{\rm i} k_x \\
{\rm i} k_x & \omega  &{\rm i} k_z &-{\rm i} k_y \\
 -\omega  &{\rm i} k_x &{\rm i} k_y &{\rm i} k_z \\
\end{array}
\right)
     \end{array}
\end{equation}
If we calculate the eigenvalues of $\mathbf{E}_0$ we obtain:
\begin{equation}\label{autovalletti}
    \lambda \, = \, \left\{-\sqrt{k_x^2+k_y^2+k_z^2}-\omega
   ,-\sqrt{k_x^2+k_y^2+k_z^2}-\omega ,\sqrt{k_x^2+k_y^2+k_z^2}-\omega
   ,\sqrt{k_x^2+k_y^2+k_z^2}-\omega \right\}
\end{equation}
This means that $\mathbf{E}_0$ has a non vanishing null space only under the on-shell condition:
\begin{equation}\label{nullospazio}
    \omega \, = \, \pm \sqrt{\mathbf{k}^2}
\end{equation}
For each choice of the sign in eq. (\ref{nullospazio}) the complex dimension of the null-space is just $2$. It remains to be seen
whether the null eigenvectors are annihilated also by $\mathbf{E}_{x,y,z}$. By explicit calculation we find that indeed they are.
Hence for each choice of the frequency $\omega$ associated with a pair of momenta $\pm \mathbf{k} \in \Lambda$ in the cubic
lattice we have $4$ real parameters. In conclusion, taking into account both frequencies the number of real parameters in a
general solution of the triholomorphic constraint associated with each momentum pair $\pm \mathbf{k}$ is 8 rather than 16 as it
is the case for a generic classical solution. In other words the triholomorphic hyperinstantons are just one half of all the
classical solutions.
\par
From the above discussion it follows that the functional integration localized on the hyperinstantons can be performed by summing
independently on each pair of lattice momenta $\pm \mathbf{k} \in \Lambda$ and for each pair integrating on the $4$ real
parameters. Furthermore if we restrict the integration to functions that are square integrable over $\mathbb{R}_+\times
\mathrm{T^3}$ we have to discard the solutions with frequency $\omega = \sqrt{\mathbf{k}^2}$ which diverge exponentially at
$\mathrm{U}\to \infty$ and keep only the solutions with the negative frequency $\omega\, =\, -\,\sqrt{\mathbf{k}^2}$. Consequently the
continuous integration for each pair of lattice momenta $\pm \mathbf{k}$ is restricted to $4$ real parameters. Equivalently we
can say that  triholomorphicity  plus square integrability reduces the number of parameters to $2$ for each momentum vector.
\subsection{The hyperinstanton functional space and octahedral orbits in the momentum lattice}
\label{organisazia}
Let us now reorganize the hyperinstanton functional space in a different order which takes advantage of the
relation between the triholomorphic constraint (\ref{lomonosov}) and Beltrami equation.
\par
The key observation is that the lattice sum $\sum_{\mathbf{k}\in \Lambda}$ can be reorganized as the following sum:
\begin{equation}\label{cobordelli}
    \sum_{\mathbf{k}\in \Lambda} \, \simeq \, \sum_{\mathcal{O}_\mathbf{n} \subset \Lambda} \,
    \sum_{\mathbf{k}\in\mathcal{O}_\mathbf{n}}
\end{equation}
where $\mathcal{O}_\mathbf{p}$ denotes a finite set of lattice momenta that form an orbit under the action of the octahedral
group $\mathrm{O}_{24} \subset \mathrm{SO(3)}$. This finite group which is isomorphic to the symmetric group $\mathrm{S_4}$ is
the point-group of the cubic lattice, namely it is the discrete subgroup of rotations that map the lattice into itself. The
advantage in the reorganization of the lattice sum according to eq.(\ref{cobordelli}) is that the possible orbits of
$\mathrm{O_{24}}$ fall into $5$ types each type including an infinite number of identical copies labeled by increasing natural
numbers $\mathbf{n}\,=\,\{n_x,n_y,n_z\}$. For each of the five types of orbits the solution of Beltrami equation and hence of the
triholomorphic constraint has a universal form containing a fixed predetermined number of parameters. In this way the functional
integration on hyperinstantons admits the following very inspiring reorganization:
\begin{equation}\label{copiole}
    \int \, \exp\left[\int_{\mathcal{M}_4} \, \mathbf{k}^x \wedge  q_\star {\mathbf{K}}^x(t)\right]\mathcal{D}q
    \, = \, \sum_{I=1}^5 \left(\sum_{\mathbf{n}}\,\int \,
    \exp\left[\int_{\mathcal{M}_4} \, \mathbf{k}^x \wedge  q^{[I]}_\star\left(\mathfrak{m}|\mathbf{n}\right)
    {\mathbf{K}}^x(t)\right]
    \,\mathrm{d}^{r_I}\mathfrak{m}\right)
\end{equation}
where:
\begin{description}
  \item[a)] ${\mathbf{K}}^x(t) $ denote the HyperK\"ahler forms of the target space depending on their moduli $t$\footnote{As we
  discuss in the next section the orthogonal matrix $\mathrm{O}_{xy}$ can always be gauged fixed to $1$, but what remains
  are the redefinitions  of the HyperK\"ahler forms $\mathbf{K}^x$ of the target manifold in terms of a rotation in the coset
  manifold $\so(3,3)/\so(3) \times \so(3)$. See eq.(\ref{kallerotti})}
  \item[b)] $q^{[I]}_\star\left(\mathfrak{m}|\mathbf{n}\right)$ denotes the general form of a hyperinstanton solution associated
  with an octahedral orbit of type $I$ and degree $\mathbf{n}$, $\mathfrak{m}$ being its parameters.
 \item[c)] $r_I$ denotes the dimension of the parameter space of the the hyperinstanton
 $q^{[I]}_\star\left(\mathfrak{m}|\mathbf{n}\right)$. The key point is that $r_I$ depends only on the type and not
 on the degree $\mathbf{n}$.
\end{description}
\par
In the sequel we show that:
\begin{equation}\label{quadratona}
  \mathcal{A}_{top}\left[q^{[I]},t\right] \, \equiv \, \int_{\mathcal{M}_4} \, \mathbf{k}^x \wedge
  q^{[I]}_\star\left(\mathfrak{m}|\mathbf{n}\right)
    {\mathbf{K}}^x(t) \, = \, \sum_{x=1}^3 \,n_x \, \mathfrak{m}^\mathcal{I}
    \mathcal{Q}^{x|I}_{\mathcal{IJ}}(t) \mathfrak{m}^\mathcal{J}
\end{equation}
where $\mathcal{Q}^{x|I}_{\mathcal{IJ}}(t)$ denotes a triplet of square $r_I \times r_I$ matrices that depend only on the moduli
of the the HyperK\"ahler forms and are the same for each orbit type, independently from the degree. In this way the functional
integration on hyperinstantons reduces to just five  gaussian integrations on continuous parameters whose results depend on
integer numbers and have to be summed over them.
\par
This strategy is effective since,  thanks to the previous results obtained by two of us in \cite{Fre:2015mla} and
\cite{Fre:2015xaa}, we already know the dimensions and the precise form of the parameter spaces for the solutions of the Beltrami
equation which are associated with each octahedral orbit $\mathcal{O}_{\mathbf{n}}$  in the  momentum lattice $\Lambda_{cubic}$.
On the other hand, as we have shown above, every solution of the Beltrami equation can be mapped into a triholomorphic
hyperinstanton.
\begin{figure}[!hbt]
\begin{center}
\iffigs
\includegraphics[height=70mm]{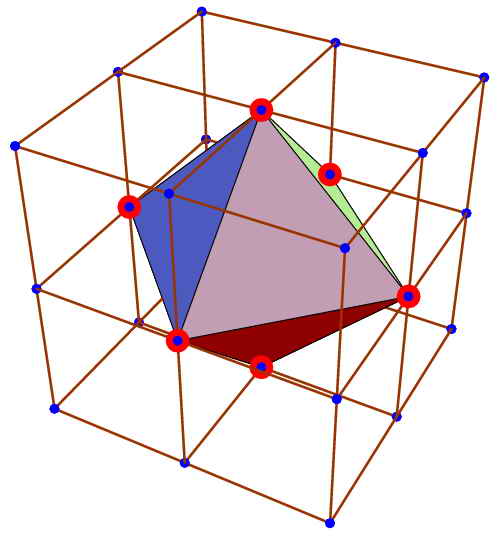}
\else
\end{center}
 \fi
\caption{\it The momenta in the cubic lattice
 forming an orbit of length $6$ under the octahedral group are of the form $\{\pm n,0,0\}$, $\{0,\pm n,0\}$,
 $\{0,0,\pm n,\}$ and correspond to the vertices of a regular octahedron.}
 \label{ottavertici}
 \iffigs
 \hskip 1cm \unitlength=1.1mm
 \end{center}
  \fi
\end{figure}
\par
Hence, as we just anticipated, there are five type of octahedral orbits and as many types of hyperinstanton spaces:
\begin{description}
  \item[1)] The octahedral orbits $\mathcal{O}_{\mathbf{n,0,0}}$ of length 6 formed by the momentum vectors
  $\{\mathbf{\pm n,0,0}\}$, $\{\mathbf{0,\pm n,0}\}$, $\{\mathbf{0,0,\pm n}\}$, where $\mathbf{n}\in \mathbb{Z}$.
  The number of parameters in the
  solution of the Beltrami equation is $6$ and the
  corresponding hyperinstanton space (\ref{generasolata}) has therefore dimension 12. The eigenvalue of the
  Laplace-Beltrami operator is:
  \begin{equation}\label{6ormu}
    \mu \, = \, \pi \, \mathbf{n}
  \end{equation}
  \item[2)] The octahedral orbits $\mathcal{O}_{\mathbf{n,n,0}}$ of length 12 formed by the momentum vectors
  $\{\mathbf{\pm n,\pm n,0}\}$, $\{\mathbf{0,\pm n,\pm n}\}$, $\{\mathbf{\pm n ,0,\pm n}\}$, where $\mathbf{n}\in \mathbb{Z}$.
  The number of parameters in the
  solution of the Beltrami equation is $12$ and the
  corresponding hyperinstanton space (\ref{generasolata}) has therefore dimension 24.
  The eigenvalue of the
  Laplace-Beltrami operator is:
  \begin{equation}\label{12ormu}
    \mu \, = \, \pi \, \sqrt{2} \,\mathbf{n}
  \end{equation}
\item[3)] The octahedral orbits $\mathcal{O}_{\mathbf{n,n,n}}$ of length 8 formed by the momentum vectors
  $\{\mathbf{\pm n,\pm n,\pm n}\}$ where $\mathbf{n}\in \mathbb{Z}$. The number of parameters in the
  solution of the Beltrami equation is $8$ and the
  corresponding hyperinstanton space (\ref{generasolata}) has therefore dimension 16.
  The eigenvalue of the
  Laplace-Beltrami operator is:
  \begin{equation}\label{12ormu}
    \mu \, = \, \pi \, \sqrt{3} \,\mathbf{n}
  \end{equation}
  \item[4)] The octahedral orbits $\mathcal{O}_{\mathbf{n,n,m}}$ of length 24 formed by the momentum vectors
  $\{\mathbf{\pm n,\pm n,\pm m}\}$, $\{\mathbf{\pm m,\pm n,\pm n}\}$, $\{\mathbf{\pm n ,\pm m,\pm n}\}$,
  where $\mathbf{n} \ne \mathbf{m}\in \mathbb{Z}$.
  The number of parameters in the
  solution of the Beltrami equation is $24$ and the
  corresponding hyperinstanton space (\ref{generasolata}) has therefore dimension 48.
  The eigenvalue of the
  Laplace-Beltrami operator is:
  \begin{equation}\label{24ormu}
    \mu \, = \, \pi \, \sqrt{ \,2\, \mathbf{n}^2 \, + \, \mathbf{m}^2}
  \end{equation}
  \item[5)] The octahedral orbits $\mathcal{O}_{\mathbf{n,m,r}}$ of length 48 formed by the momentum vectors
  $\{\mathbf{\pm n,\pm m,\pm r}\}$, $\{\mathbf{\pm m,\pm n,\pm r}\}$, $\{\mathbf{\pm n,\pm r,\pm m}\}$,
  $\{\mathbf{\pm r,\pm m,\pm n}\}$,$\{\mathbf{\pm m,\pm r,\pm n}\}$,
  $\{\mathbf{\pm r,\pm n,\pm m}\}$
  where $\mathbf{n} \ne \mathbf{m}\ne \mathbf{r}\in \mathbb{Z}$.
  The number of parameters in the
  solution of the Beltrami equation is $48$ and the
  corresponding hyperinstanton space (\ref{generasolata}) has therefore dimension 96.
  The eigenvalue of the
  Laplace-Beltrami operator is:
  \begin{equation}\label{24ormu}
    \mu \, = \, \pi \, \sqrt{ \, \mathbf{n}^2 \, + \, \mathbf{m}^2+ \, \mathbf{r}^2}
  \end{equation}
\end{description}
\begin{figure}[!hbt]
\begin{center}
\iffigs
\includegraphics[height=70mm]{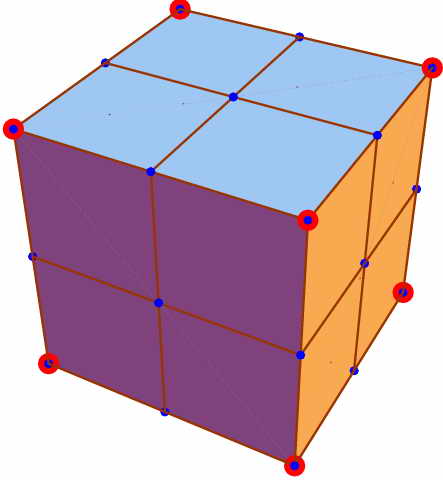}
\else
\end{center}
 \fi
\caption{\it The momenta in the cubic lattice
 forming an orbit of length $8$ under the octahedral group are of the form $\{\pm n,\pm n,\pm n\}$,
  and correspond to the vertices of a regular cube.}
 \label{cubamezzi}
 \iffigs
 \hskip 1cm \unitlength=1.1mm
 \end{center}
  \fi
\end{figure}
\begin{figure}[!hbt]
\begin{center}
\iffigs
\includegraphics[height=70mm]{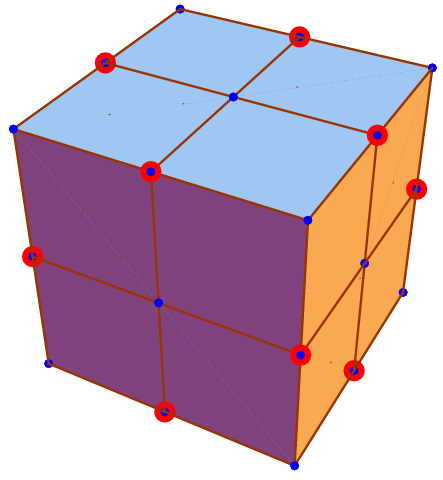}
\else
\end{center}
 \fi
\caption{\it The momenta in the cubic lattice
 forming an orbit of length $12$ under the octahedral group are of the form $\{\pm n,\pm n,0\}$,$\{0,\pm n,\pm n\}$,
 $\{\pm n,0, \pm n\}$
  and correspond to mid-points of the edges of a regular cube.}
 \label{cubavertici}
 \iffigs
 \hskip 1cm \unitlength=1.1mm
 \end{center}
  \fi
\end{figure}
According to this classification we immediately have the dimensions of the parameter spaces for the  various types of orbits:
\begin{eqnarray}
  r_{\mathbf{n,0,0}} &=& 12 \nonumber \\
 r_{\mathbf{n,n,0}} &=& 24 \nonumber \\
  r_{\mathbf{n,n,n}} &=& 16 \nonumber \\
 r_{\mathbf{n,n,m}} &=& 48 \nonumber \\
 r_{\mathbf{n,m,r}} &=& 96 \label{anomalies}
\end{eqnarray}
\subsection{The role of the Universal Classifying Group $\mathrm{\mathrm{G_{1536}}}$} \label{universalone} In
\cite{Fre:2015xaa},\cite{Fre:2015mla}, inspired by the space group  constructions of crystallography and by Frobenius
congruences,  two of us introduced an extension of the octahedral group by means of translations quantized in units of
$\frac{1}{4}$. In each direction and modulo integers there are just four translations $0, \, \ft 14, \, \ft 12, \, \ft 34$ so
that the translation subgroup reduces to $\mathbb{Z}_4 \, \otimes\,\mathbb{Z}_4\, \otimes \, \mathbb{Z}_4$  that has a total of
$64$ elements. In this way we singled out a discrete group $\mathrm{G_{1536}}$ of order $24 \times 64 \,  =  \, 1536$,  which is
simply the semidirect product of the point group $\mathrm{O_{24}}$ with $\mathbb{Z}_4 \, \otimes\,\mathbb{Z}_4\, \otimes \,
\mathbb{Z}_4$:
\begin{equation}\label{1536defino}
     \mathrm{\mathrm{G_{1536}}} \, \simeq \, \mathrm{O_{24}} \,
    \ltimes \, \left (\mathbb{Z}_4 \, \otimes\,\mathbb{Z}_4\, \otimes \, \mathbb{Z}_4\right)
\end{equation}
We named $\mathrm{\mathrm{G_{1536}}}$ the universal classifying
group of the cubic lattice.
\begin{figure}[!hbt]
\begin{center}
\iffigs
\includegraphics[height=70mm]{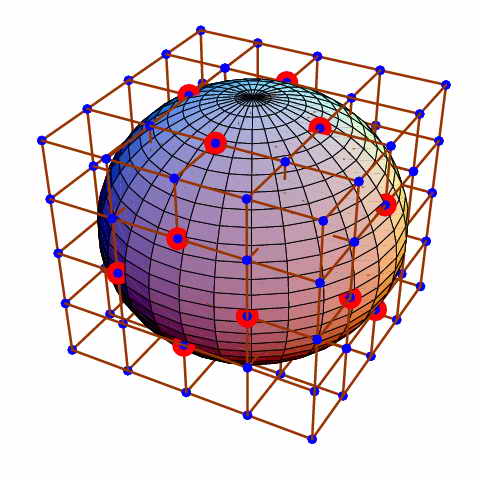}
\else
\end{center}
 \fi
\caption{\it  A view of an  orbit of length 24 in the cubic lattice: the lattice points are of the form $\{\pm a, \pm a,\pm b\}
$, $\{\pm a, \pm b,\pm a\}$,$\{\pm b, \pm a,\pm a\}$ and intersect the sphere of radius $r^2 \, = \, 2 a^2+b^2$}
\label{orbita24PS}
 \iffigs
 \hskip 1cm \unitlength=1.1mm
 \end{center}
  \fi
\end{figure}
For all details relative to such a group we refer the reader to the two papers \cite{Fre:2015xaa}, \cite{Fre:2015mla}. What is
relevant to us here is that $\mathrm{\mathrm{G_{1536}}}$ is certainly a global symmetry group of the topological sigma-model  and
that the parameter spaces of hyperinstantons decompose into irreducible representations of $\mathrm{\mathrm{G_{1536}}}$.
\par
As shown in \cite{Fre:2015xaa},\cite{Fre:2015mla} the group $\mathrm{\mathrm{G_{1536}}}$  has 37 conjugacy
classes whose populations are distributed as follows:
\begin{description}
  \item[1)]  2 classes of length 1
  \item[2)]  2  classes of length 3
  \item[3)]  2 classes of  length 6
  \item[4)] 1 class of length 8
  \item[5)] 7 classes of length 12
  \item[6)] 4 classes of length 24
  \item[7)] 13 classes of length 48
  \item[8)] 2 classes of length 96
  \item[9)] 4 classes of length 128
\end{description}
It follows that there are $37$  irreducible representations whose construction was performed in \cite{Fre:2015xaa} relying
on the method of induced representations. The $37$ irreps are distributed according to the following pattern:
\begin{description}
  \item[a)] 4 irreps of dimension $1$, namely $\mathrm{D}_1,\dots,\mathrm{D}_4$
  \item[b)] 2 irreps of dimension $2$, namely $\mathrm{D}_5,\dots,\mathrm{D}_6$
  \item[c)] 12 irreps of dimension $3$, namely $\mathrm{D}_6,\dots,\mathrm{D}_{18}$
  \item[d)] 10 irreps of dimension $6$, namely $\mathrm{D}_7,\dots,\mathrm{D}_{28}$
  \item[e)] 3 irreps of dimension $8$, namely $\mathrm{D}_{29},\dots,\mathrm{D}_{31}$
  \item[f)] 6 irreps of dimension $12$, namely $\mathrm{D}_{32},\dots,\mathrm{D}_{37}$
\end{description}
In \cite{Fre:2015xaa} two of us thoroughly discussed the decomposition of the parameter-spaces of Beltrami vector fields
(\textit{i.e.} solutions of the Beltrami equation) into irreps of $\mathrm{\mathrm{G_{1536}}}$. Due to the relation between the
hyperinstantons and Beltrami fields explained in the previous section such a  decomposition extends canonically to the
hyperinstanton parameter-spaces. In this way, we can make a third reorganization of the functional integral (\ref{copiole}) as
 a discrete sum on the $37$ irreducible representations plus a sum on their multiplicities and
 a continuous integral on the parameters pertaining to any such a representation.
 \par
From the above list, if we were to do so  we see that the maximal dimension of an irreducible moduli-space is at most 12 and not
96 as displayed in eq.(\ref{anomalies}) for reducible moduli-spaces.
\par
In particular in the present paper we restrict our attention to consider the parameter-spaces  associated with the smallest
momentum space octahedral orbits $\mathcal{O}_{\{\mathbf{n,0,0}\}}$ of length 6. As explained in \cite{Fre:2015xaa}, there are
four classes of momentum vectors yielding orbits of length 6: $\{\mathbf{n,0,0} \}$}. The decomposition into irreducible
representations of the 12-dimensional moduli-spaces $ \mathfrak{M}^{[12]}_{\{\mathbf{n ,0,0}\}}$ is displayed below\footnote{The
results mentioned here and in all the sequel of this paper have been obtained by means of MATHEMATICA codes built on the basis of
the corpus of MATHEMATICA codes utilized to derive the systematics exposed in \cite{Fre:2015xaa}.}.
 \begin{enumerate}
   \item Class $\mathbf{k }\, = \, \{\mathbf{1+4 p ,0,0}\}$ ( where $ \mathbf{p} \in \mathbb{Z}$).
   The splitting of the 12-dimensional moduli space is the following one:
       \begin{equation}\label{recogno1}
        \mathfrak{M}^{[12]}_{\{\mathbf{1+4 p ,0,0}\}} \, = \,
        \underbrace{\mathrm{D_{23}}\left[\mathrm{G_{1536}},6\right]}_{Beltrami} \oplus \mathrm{D_{22}}\left[\mathrm{G_{1536}},6\right]
       \end{equation}
    \item Class $\mathbf{k }\, = \, \{\mathbf{2+4 p ,0,0}\}$ ( where $ \mathbf{p} \in \mathbb{Z}$).
    The splitting of the 12-dimensional moduli space is the following one:
       \begin{equation}\label{recogno2}
        \mathfrak{M}^{[12]}_{\{\mathbf{2+4 p ,0,0}\}} \, = \,
        \underbrace{\mathrm{D_{19}}\left[\mathrm{G_{1536}},6\right]}_{Beltrami}
        \oplus \mathrm{D_{9}}\left[\mathrm{G_{1536}},3\right]\oplus \mathrm{D_{11}}\left[\mathrm{G_{1536}},3\right]
       \end{equation}
   \item Class $\mathbf{k }\, = \, \{\mathbf{3+4 p ,0,0}\}$ ( where $ \mathbf{p} \in \mathbb{Z}$).
   The splitting of the 12-dimensional moduli space is the following one:
       \begin{equation}\label{recogno3}
        \mathfrak{M}^{[12]}_{\{\mathbf{3+4 p ,0,0}\}} \, = \, \underbrace{\mathrm{D_{24}}
        \left[\mathrm{G_{1536}},6\right]}_{Beltrami} \oplus \mathrm{D_{22}}\left[\mathrm{G_{1536}},6\right]
       \end{equation}
   \item Class $\mathbf{k }\, = \, \{\mathbf{0+4 p ,0,0}\}$ ( where $ \mathbf{p} \in \mathbb{Z}$).
   The splitting of the 12-dimensional moduli space is the following one:
       \begin{equation}\label{recogno4}
        \mathfrak{M}^{[12]}_{\{\mathbf{0+4 p ,0,0}\}} \, = \, \underbrace{\mathrm{D_{7}}
        \left[\mathrm{G_{1536}},3\right] \oplus \mathrm{D_{8}}\left[\mathrm{G_{1536}},3\right]}_{Beltrami}\oplus\mathrm{D_{8}}
        \left[\mathrm{G_{1536}},3\right] \oplus \mathrm{D_{5}}\left[\mathrm{G_{1536}},2\right]
        \oplus \mathrm{D_{1}}\left[\mathrm{G_{1536}},1\right]
       \end{equation}
 \end{enumerate}
The representations underwritten \textit{Beltrami} are those that already appear in the decomposition of the Beltrami field
$\mathbf{Y}^{[1]}$ (see \cite{Fre:2015xaa}). The other representations are those that are contributed by the function
$G(\mathbf{X})$ (see eq. (\ref{factoF})).
\section{The coupling constants in the topological action and their geometry}
\label{topolconstant}
 We observe that a HyperK\"ahler space is in particular K\"ahler and furthermore that our flat manifold is
 a Calabi-Yau two-fold: it has vanishing first Chern class of the tangent bundle $c_1(\mathcal{T}\mathbb{R}^4) \, = \,
0$.
\par
In order to appreciate the  geometrical meaning of the parameters in the Lagrangian and the new quality of the
hyperinstantons occurring in the 4D sigma-model with respect to the holomorphic instantons occurring in the 2D sigma
model, we will make a strict comparison with the classical results concerning those 2D $\sigma$-models whose target
space is selected to be the Kummer surface K3. Since this latter is also HyperK\"ahler it could be taken as target
manifold also in the  4D case and comparison with it provides a formidable instructive case.
\par
In view of that discussion of the K3 case it becomes clear what is the relevant structure of the topological coupling constants
in the lagrangian of the $\mathrm{4D}$ topological sigma-model under consideration. First of all, by means of a simple argument
we can dispose the matrix $\mathrm{O}_{xy}$ that we have so far carried through our discussion.
\par
According to the topological twist, the correct hyperinstanton equations are those where the equation is just
(\ref{lomonosov}). Yet given the above general solution (\ref{generasolata}) of such an equation, let us consider the
following linear transformation on the solution:
\begin{equation}\label{so3m}
    \hat{q}^u(\mathrm{U},\mathbf{X})\, = \,{q}^v(\mathrm{U},\mathbf{X}) \mathcal{S}_v^{\phantom{v}u} \quad ; \quad
    \mathcal{S}\, \in \, \mathrm{SO^-(3)} \subset \mathrm{SO(4)}
\end{equation}
where, by definition,$\mathrm{SO^-(3)}$ is the subgroup of $\mathrm{SO(4)}$ that rotates the antiself-dual 't Hoft
matrices among themselves. $\mathrm{SO^-(3)}$ is generated by the very triplet of antiself-dual matrices $J^{-|x}$ and
acts as follows:
\begin{equation}\label{azionecorta}
    \mathcal{S}^T \,J^{-|x} \, \mathcal{S} \, = \, \left(\mathrm{O}\left[\mathcal{S}\right]\right)^{xy} \, J^{-|y}
\end{equation}
where we have denoted by $\mathrm{O}\left[\mathcal{S}\right]$ the homomorphism:
\begin{equation}\label{conterrus}
\mathrm{SO(4)}  \, \subset \,  \mathrm{SO^-(3)} \, \stackrel{\mathrm{O}}{\Longrightarrow} \, \mathrm{SO(3)} \, = \, 3
\times 3 \mbox{ matrices}
\end{equation}
 From equation (\ref{azionecorta}) it follows that if ${q}^u(\mathrm{U},\mathbf{X})$ is triholomorphic with
respect to $\mathrm{O} \, = \, 1$, than $\hat{q}^u(\mathrm{U},\mathbf{X})$ is triholomorphic with respect
$\mathrm{O}[\mathcal{S}]$. This shows that given (\ref{generasolata}) we easily obtain the general solution of the
hyperinstanton equation for any matrix $\mathrm{O}\in \mathrm{SO(3)}$. On the other hand the transformation
(\ref{so3m}) is just a diffeomorphism of the target space, which is even an isometry. Hence the new hyperinstanton
obtained by composing the original map with $\mathcal{S}$ is to be identified with the earlier one and does not
correspond to a truly new solution. Fixing $\mathrm{O}\, = \, \mathbf{1}_{3\times 3}$  is just a convenient
gauge-fixing of this symmetry. On the other hand we can write:
\begin{equation}\label{kallerotti}
    \mathbf{K}^{x} \, = \, \mathbb{L}(t)^{{x}J} \, \alpha_J \quad ; \quad J\, = \, 1,2,3,\dot{1},\dot{2},\dot{3}
    \quad ; \quad x=1,2,3
\end{equation}
where $\alpha_J$ denotes the basis of harmonic two-forms defined in eq.(\ref{arappo}) and  the $6\times 6$matrix:
\begin{equation}\label{cosettus}
    \mathbb{L}(t)^{IJ} \, \in \, \frac{\mathrm{SO(3,3)}}{\mathrm{SO(3)\times SO(3)}} / \mathrm{O(3,3,\mathbb{Z})}
\end{equation}
is a coset representative depending on $9$ parameters $t^\alpha$ that can be regarded as the true topological coupling constants
of the topological sigma model under consideration. Let us then recall the local isomorphism $\so(3,3) \simeq
\slal(4,\mathbb{R})$ discussed in appendix \ref{isomorfo}. As we show there, the deformations of the HyperK\"ahler 2-forms
encoded in eq.(\ref{kallerotti}) are equivalent to deformations of the flat metric of $\mathbb{R}^4$. We can equivalently
consider the space of flat metrics or the space of flat HyperK\"ahler 2-forms on $\mathbb{R}^4$. On the other hand
eq.(\ref{kallerotti}) has a very simple interpretation. Just as in the case of the an algebraic K3 model in $d=2$ , the
topological coupling constants were, for the $A$-twisted model the parameters of the K\"ahler class, for the B-twisted model the
parameters of the complex structure, in the case under consideration, the topological coupling constants correspond to the choice
of an entire HyperK\"ahler structure, whose space is indeed that spelled out in eq.(\ref{cosettus}).
\par
For $\mathbb{L}(t)^{IJ}(t)$ one can choose various types of parameterizations, a simple one being the solvable parametrization \cite{solvabli},
discussed in Appendix \ref{fractional}. Another parametrization which is quite convenient to our goals is the classical
off-diagonal one also recalled in the  appendix     \ref{offodiag}. Let us also observe that provided we are able to determine
them analytically no one prevents us to consider hyperinstantons in the topological sigma model where the target space is K3,
rather then $\mathbb{R}^4$. In that case eq.(\ref{kallerotti}) would apply equally well with:
\begin{equation}\label{cosettusK3}
    \mathbb{L}(t)^{IJ} \, \in \, \frac{\mathrm{SO(3,19)}}{\mathrm{SO(3)\times SO(19)}} / \mathrm{O(3,19,\mathbb{Z})}
\end{equation}
instead of eq.(\ref{cosettus}).
\section{Moduli space of the hyperinstantons, boundary conditions and the BRST complex}
We come next to  the conceptual questions left open in section \ref{topocurvo} about boundary conditions and BRST invariance. In
particular we address the discussion of the topological action (\ref{paronzo}) and the question about its role as classifier of
the homotopy classes. The first key observation is the following one. When the target space is compact with a non trivial
geometry,  like $\mathrm{K3}$, the triholomorphic constraint is a set of non-linear differential equations and multiplication of
a solution by a constant does not produce a new solution. In our case where the target space has a flat geometry the
triholomorphic constraint is linear and indeed, as we have shown, equivalent to Beltrami equation. This implies that if we
multiply a solution by an arbitrary constant $\lambda \in \mathbb{R}$, the result is still a solution. Correspondingly the
topological action is rescaled by $\lambda^2$. Clearly such a scale deformation of the solutions is not an interesting one from
the point of view of the homotopy of the map. We are interested in those deformations that keep the topological action invariant
at a fixed scale. Then by definition, the moduli space of the hyperinstantons corresponding to deformations within the same
homotopy class are  those that keep the topological action invariant. This choice  allows to obtain BRST invariance of the
action.
\par
In this section  we  illustrate this issue utilizing the concrete example of the hyperinstantons in the shortest octahedral orbit
$\mathcal{O}_{\mathbf{n,0,0}}$ which, as anticipated above, form a 12-parameters class. Explicitly the form of this map is the
following one:
\begin{eqnarray}
  q^0 &=& e^{-2\pi \, n\,\mathrm{U}}\left[-\mathfrak{m}_{10} \sin (2 \pi  n \mathrm{X})+\mathfrak{m}_7 \cos (2 \pi  n \mathrm{X})-\mathfrak{m}_{11}
  \sin (2 \pi  n \mathrm{Y})\right.\nonumber\\
  &&\left.+\mathfrak{m}_8\cos (2 \pi  n \mathrm{Y})-\mathfrak{m}_{12} \sin (2 \pi  n \mathrm{Z})
  +\mathfrak{m}_9 \cos (2 \pi  n \mathrm{Z})\right]\nonumber \\
  q^x &=& e^{-2\pi \, n\,\mathrm{U}}\left[\mathfrak{m}_7 \sin (2 \pi  n \mathrm{X})+\mathfrak{m}_{10}
  \cos (2 \pi  n \mathrm{X})-\mathfrak{m}_5 \sin (2 \pi  n \mathrm{Y})\right.\nonumber\\
  &&\left.+\mathfrak{m}_2
   \cos (2 \pi  n \mathrm{Y})+\mathfrak{m}_3 \sin (2 \pi  n \mathrm{Z})+\mathfrak{m}_1 \cos (2 \pi  n \mathrm{Z})\right]\nonumber \\
  q^y &=& e^{-2\pi \, n\, \mathrm{U}}\left[\mathfrak{m}_6 \sin (2 \pi  n \mathrm{X})+\mathfrak{m}_4 \cos (2 \pi  n \mathrm{X})
  +\mathfrak{m}_8 \sin (2 \pi  n \mathrm{Y})\right.\nonumber\\
  &&\left.+\mathfrak{m}_{11}
   \cos (2 \pi  n \mathrm{Y})+\mathfrak{m}_1 (-\sin (2 \pi  n \mathrm{Z}))+\mathfrak{m}_3 \cos (2 \pi  n \mathrm{Z})\right]\nonumber \\
  q^z &=& e^{-2\pi \, n\,\mathrm{U}}\left[-\mathfrak{m}_4 \sin (2 \pi  n \mathrm{X})+\mathfrak{m}_6 \cos (2 \pi  n \mathrm{X})
  +\mathfrak{m}_2 \sin (2 \pi  n \mathrm{Y})\right.\nonumber\\
  &&\left.+\mathfrak{m}_5 \cos
   (2 \pi  n \mathrm{Y})+\mathfrak{m}_9 \sin (2 \pi  n \mathrm{Z})+\mathfrak{m}_{12} \cos (2 \pi  n \mathrm{Z})\right]
\end{eqnarray}
In order to calculate the topological action functional(\ref{quadratona}), we begin by defining the deformed HyperK\"ahler forms
of the target manifold. We find it convenient to utilize the off-diagonal parametrization of the coset
$\mathbb{H}_{\mathrm{3,3}}$ displayed in eq.(\ref{posaidone}). Naming $\Gamma \, = \, \sqrt{1+\Psi\,\Psi^T}$, the topological
action has the form
\begin{equation}\label{coreutone}
  \mathcal{A}_{top}\left[q \right]\, = \, \Gamma_{xy} \, \mathcal{Z}_{xy} \, + \, \Psi_{x\dot{y}} \mathcal{Z}_{x\dot{y}}
\end{equation}
where we have defined:
\begin{equation}\label{Zkernel}
    \mathcal{Z}_{{I}{J}} \, = \, \int_{\mathbb{R}_+\times \mathrm{T^3}} \,
    \alpha_{I} \wedge  q_\star[
    {\alpha}_{J}]
\end{equation}
Performing the explicit integrations, which are all convergent due to the exponential factor $ e^{-2\pi \, n\,\mathrm{U}}$, we
obtain:
\begin{equation}\label{fortedeimarmi}
    \mathcal{A}_{top}\left[q \right] \, = \,4\pi \, n \, \ft 12 \, \mathfrak{m}^\mathcal{I} \, \mathcal{Q}_{\mathcal{IJ}} \,
    \mathfrak{m}^\mathcal{J}
\end{equation}
where the $12\times 12$ matrix $\mathcal{Q}$ has the  explicit form given in eq.(\ref{fanciullus}). The 12 eigenvalues of
$\mathcal{Q}$ are the following ones:
\begin{eqnarray}
  \lambda &=& 4 \pi \, n \left\{\gamma _{1,1}-\sqrt{\psi _{1,1}^2+\psi _{1,2}^2+\psi _{1,3}^2},\,\gamma
   _{1,1}-\sqrt{\psi _{1,1}^2+\psi _{1,2}^2+\psi _{1,3}^2},\right. \nonumber\\ && \left.
   \gamma_{1,1}+\sqrt{\psi _{1,1}^2+\psi _{1,2}^2+\psi _{1,3}^2},\gamma
   _{1,1}+\sqrt{\psi _{1,1}^2+\psi _{1,2}^2+\psi _{1,3}^2}, \right. \nonumber\\ && \left.
   \gamma_{2,2}-\sqrt{\psi _{2,1}^2+\psi _{2,2}^2+\psi _{2,3}^2},\,
   \gamma_{2,2}-\sqrt{\psi _{2,1}^2+\psi _{2,2}^2+\psi _{2,3}^2},\right. \nonumber\\ && \left.
   \gamma_{2,2}+\sqrt{\psi _{2,1}^2+\psi _{2,2}^2+\psi _{2,3}^2},\,
   \gamma_{2,2}+\sqrt{\psi _{2,1}^2+\psi _{2,2}^2+\psi _{2,3}^2},\right. \nonumber\\ && \left.
   \gamma_{3,3}-\sqrt{\psi _{3,1}^2+\psi _{3,2}^2+\psi _{3,3}^2},\,
   \gamma_{3,3}-\sqrt{\psi _{3,1}^2+\psi _{3,2}^2+\psi _{3,3}^2},\right. \nonumber\\ && \left.
   \gamma_{3,3}+\sqrt{\psi _{3,1}^2+\psi _{3,2}^2+\psi _{3,3}^2},\,
   \gamma_{3,3}+\sqrt{\psi _{3,1}^2+\psi _{3,2}^2+\psi _{3,3}^2}\right\}
\end{eqnarray}
having denoted by $\gamma_{ij}$ and $\psi_{ij}$ the matrix elements of  the matrices $\Gamma$ and $\Psi$, respectively. The
matrix $\mathcal{Q}$ can be put into diagonal form by an orthogonal similarity transformation:
\begin{equation}\label{singhiozzo}
    \mathfrak{C}^T \, \mathcal{Q} \, \mathfrak{C} \, = {\rm diag}\, (\lambda_1, \dots, \lambda_{12})
\end{equation}
where the orthogonal matrix $\mathfrak{C}$ is easy to calculate but too large to be displayed.  Going back to equation
\ref{fortedeimarmi}  we can redefine the hyperinstanton parameters as follows:
\begin{equation}\label{miny}
    \mathfrak{m} \, = \,\mathfrak{C} \, \mathfrak{p}
\end{equation}
and the topological action  becomes:
\begin{equation}\label{ferocissimo}
    \mathcal{A}_{top}\left[q \right] \, = \, 2 \, \pi \, n \, \sum_{\mathcal{I}=1}^{12} \,
    \lambda_\mathcal{I} \, \mathfrak{p}_\mathcal{I}^2
\end{equation}
Since the matrix $\mathfrak{C}$ is orthogonal, the integration measure remains up to this stage unchanged $\mathrm{d}^{12}
\mathfrak{m} \, = \,\mathrm{d}^{12} \mathfrak{p}$. The next step consists of rescaling the variables $\mathfrak{p}_\mathcal{I}$,
by setting:
\begin{equation}\label{frillus}
    \mathfrak{p}_\mathcal{I} \, = \, \frac{1}{\sqrt{ \lambda_\mathcal{I}}} \mathfrak{s}_\mathcal{I}
\end{equation}
Obviously the same replacements (\ref{miny},\ref{frillus}) have to be done in the integrand $(\dots|\mathfrak{p},n)$, whatever it
is. As a final result of this procedure  the contribution of the orbits of type $\{\mathbf{n,0,0}\}$ to the functional integral
is turned into:
\begin{equation}\label{integrullo}
 \mbox{sum on these hyperinst. of }(\dots) \, = \,
 \sum_{n=1}^\infty \, \frac{1}{ \sqrt{\mbox{det}\mathcal{Q}}} \,\int \,
 \mathrm{d}^{12} \mathfrak{s} \, \exp\left[ - 2\pi \, n\, \sum_{\mathcal{I}=1}^{12} \, \mathfrak{s}_\mathcal{I}^2 \right]\times
  (\dots|\mathfrak{s},n,\mathfrak{C},\lambda)
\end{equation}
where we have carefully specified that after the variable transformation the integrand obtains dependence on  the eigenvalues
$\lambda$ and the entries of the orthogonal matrix $\mathfrak{C}$. If the integrand did not depend on the degree $n$ we might
immediately perform the summation on the integers, yet this is not the case because the integrand has also an $n$-dependence
which is not fixed a priori.
\par
Next we have to take into account the observation we made about the scale of the map. Such a scale is encoded in the moduli
$\mathfrak{s}_\mathcal{I}$. We set:
\begin{equation}\label{cavolfiorecotto}
    \mathfrak{s}_\mathcal{I} \, = \, \rho \, \mathfrak{u}_I \quad ; \quad \sum_{\mathcal{I}=1}^{12} \, \mathfrak{u}_\mathcal{I}^2
    \, = \, 1
\end{equation}
The constrained $\mathrm{u}_I$ span a sphere $\mathbb{S}^{11}$ so that  we can write:
\begin{eqnarray}\label{integrullo}
 \mbox{sum on these hyp. of }(\dots) & = &
 \sum_{n=1}^\infty \, \frac{1}{ \sqrt{\mbox{det}\mathcal{Q}}} \,\int_0^\infty \, d\rho \rho^{11}
   \, \exp\left[ - 2\pi \, n\, \rho^2 \right]\times \int \mathrm{d}\Omega^{11}(\chi)
  (\dots|\chi ,n,\mathfrak{C},\lambda)\nonumber\\
  &=& \frac{15}{16 \pi ^6}\sum_{n=1}^\infty \, \frac{1}{ n^6 }\times \int \mathrm{d}\Omega^{11}(\chi)
  (\dots|\chi ,n,\mathfrak{C},\lambda)
\end{eqnarray}
where the integrand has been supposed to depend on the scale $\rho$ as $\rho^{11}$. This is the same as assuming that the true
hyperinstantons moduli are only the angles of $\mathbb{S}^{11}$. This assumption is formally justified in the next section.
\subsection{Boundary conditions on the ghost fields}
So far we have discussed the bosonic fields. The original $\mathcal{N}=2$ $\sigma$-models contains also the fermion fields that,
after the twist, become the ghost and the antighosts. The functional integral has to be done also on these fields and we have to
work out their appropriate boundary conditions in order to define their functional space. To this effect let us consider the
boundary term presented in equations (\ref{discesona}) and (\ref{farmalucco}). The explicit form of the three-form
$\mathfrak{I}^{3,1}$ reduced to the boundary is the following one:
\begin{equation}\label{fascinoso}
    \mathfrak{I}^{3,1} \, = \, \mbox{const} \,\times \, K^{-|x}_{\mathbf{ij}} \,K^{x}_{uv}(t)\,c^v(\mathbf{X})\,
    \partial_\mathbf{k}q^u(\mathbf{X}|\mathbf{m})\, \mathrm{d}\mathrm{X}^\mathbf{i}\wedge \mathrm{d}\mathrm{X}^\mathbf{j}
    \wedge \mathrm{d}\mathrm{X}^\mathbf{k}
\end{equation}
where the coordinate $\mathrm{U}$ has been fixed to zero, $c^v(\mathbf{X})$ is the ghost field on the $\mathrm{T^3}$ boundary,
the indices $\mathbf{i,j,k}$ run on the three values $x,y,z$ while the indices $u,v$ run on the 4 values $0,x,y,z$.
BRST-invariance of the topological action requires that the integral of $\mathfrak{I}^{3,1}$ on the boundary should be zero. This
is a clear-cut functional constraint on the ghost fields. At first sight it seems a rather complicated condition, yet in the
semiclassical approximation utilized in topological field theories it becomes rather simple.
\par
We just follow the discussion presented in the book \cite{n2wonder} at page 316, eq.(7.7.60) and following ones. In the
background of a hyperinstanton $q^u(\mathrm{U},\mathbf{X}|\mathfrak{m})$, the ghost field can be expanded as follows:
\begin{equation}\label{casinaro}
    c^u(\mathrm{U},\mathbf{X})\, = \, \sum_{\alpha=1}^\Delta \theta^\alpha \, \delta q^u_\alpha(\mathrm{U},\mathbf{X}|\mathfrak{m}))
    \, + \, \delta c^u(\mathrm{U},\mathbf{X})
\end{equation}
where $\delta c^u$ are higher modes (not satisfying the triholomorphic constraint) while $\delta
q^u_\alpha(\mathrm{U},\mathbf{X},|\mathfrak{m}))$ are a basis of zero-modes, namely a set solutions of the triholomorphic
equation that furthermore satisfy the boundary condition (\ref{fascinoso}) in the background of the hyperinstanton
$q^u(\mathrm{U},\mathbf{X},|\mathfrak{m}))$. Because of the linearity of the triholomorphic constraint we can write:
\begin{equation}\label{fluttuazie}
    \delta q^u_\alpha(\mathrm{U},\mathbf{X},|\mathfrak{m}) \, = \, \kappa_\alpha^{\mathcal{I}} \frac{\partial}{\partial
    \mathfrak{m}^{\mathcal{I}}} q^u_\alpha(\mathrm{U},\mathbf{X},|\mathfrak{m})
\end{equation}
where $\kappa^{\mathcal{I}}_\alpha$ is a set of $\Delta$ 12-vectors such that the constraint (\ref{fascinoso}) is satisfied by
the ghost zero mode. Inserting the expression (\ref{fluttuazie}) in the condition (\ref{fascinoso}) and performing the integrals
we obtain:
\begin{equation}\label{condizionas}
    \mathfrak{m}^{\mathcal{I}} \, \mathcal{Q}_{\mathcal{IJ}} \, \kappa^{\mathcal{J}}_\alpha \, = \, 0
\end{equation}
The number of solutions of the algebraic equation (\ref{condizionas}) is obviously 11 and the best way of solving the equation is
by setting:
\begin{equation}\label{ilperipeo}
    \kappa_\alpha \, = \, \rho \, \mathfrak{C}\, \mathcal{D} \, \frac{\partial}{\partial \chi^\alpha} \mathfrak{u}(\chi)
\end{equation}
where:
\begin{equation}\label{Dmatru}
    \mathcal{D}  \, = \,  {\rm diag}\left(\frac{1}{\sqrt{\lambda_1}}, \dots, \frac{1}{\sqrt{\lambda_{12}}} \right)
\end{equation}
and $\frac{\partial}{\partial \chi^\alpha}$ is the derivative with respect to the Euler angles parameterizing the
$\mathbb{S}^{11}$ sphere. As explained in \cite{n2wonder} the Berezin integration on the fermionic variables $\theta^\alpha$
replaces them with the associated differential $\mathrm{d}\chi^\alpha$ and the final outcome of the story is that the appropriate
ghost fields that satisfy the boundary conditions and give BRST invariance of the action are:
\begin{eqnarray}\label{congressoditopi}
    c^u & = & \mathrm{d}\chi^\alpha \,\frac{\partial}{\partial
    \chi^\alpha}q^u(\mathrm{U},\mathbf{X}|\mathfrak{m}(\rho,\chi))\nonumber\\
    & = &\rho \,\mathrm{d}\chi^\alpha \,\frac{\partial}{\partial
    \chi^\alpha}q^u(\mathrm{U},\mathbf{X}|\mathfrak{m}(1,\chi))
\end{eqnarray}
The topological observables are linear functions of the ghost fields and hence $1$-forms on the sphere $\mathbb{S}^{11}$. The only
possible correlator is the top-form of degree eleven and the assumption on the dependence $\rho^{11}$ of the integrand is
demonstrated.

\section{Summary and Conclusions}
We reconsidered  triholomoprhic maps from a four dimensional HyperK\"ahler space
to a $4$-dimensional target HK space motivated by the intriguing correspondence between the Beltrami equation,
the triholomorphic maps and the hyperinstantons. We used such a relation to count the number of solutions (and
their deformations or, in other words, their moduli) by means of the localization of the path integral around triholomorphic maps.
We showed that in a 4d worldvolume space a twisting procedure can be performed along the ways of  the topological 2d sigma models and that it gives us some very powerful computational means to understand the (quantum) geometry of the solutions
to the Beltrami equation. This paper is the starting point for a more ambitious goal of computing the complete
range of solutions to the Beltrami equation by studying the topological correlators of  a sigma model.
\par
Combining the role of Beltrami vector fields as fluxes in $2$-brane solutions of $D=7$ supergravity with their reinterpretation as building blocks of hyperinstantons opens the perspective of relating hyperinstantons to supergravity $2$-branes which is quite challenging and will be the object of further investigations.
\newpage
\appendix
\section{Complex structure of the hyperinstantons}
\label{complessone}
 To appreciate the precise nature of  the hyperinstanton triholomorphicity equations and of the topological
action it is convenient to go to a complex basis. Although our goal is the specific choice $\mathcal{N}_{4}\, = \, \mathbb{R}^4$
for the target manifold, in the present discussion we maintain the choice of the latter arbitrary in order to emphasize the
formal structure of the hyperinstantons.
\par
We adopt the following conventions. The hyperK\"ahler structures on the base manifold and on the target manifold are respectively
chosen as follows:
\begin{equation}\label{ipercallero}
   \begin{array}{|ccc|ccc|}
   \hline
     \omega^{(2,0)} & = & \frac{\mathbf{k}^{2}+{\rm i}\mathbf{k}^1}{\sqrt{2}} &
     \Omega^{(2,0)} & = & \frac{\mathbf{K}^{2}+{\rm i}\mathbf{K}^1}{\sqrt{2}} \\
     \omega^{(0,2)} & = & \frac{\mathbf{k}^{2}-{\rm i}\mathbf{k}^1}{\sqrt{2}} &
     \Omega^{(0,2)} & = & \frac{\mathbf{K}^{2}-{\rm i}\mathbf{K}^1}{\sqrt{2}} \\
     \omega^{(1,1)} & = & \mathbf{k}^3 &
     \Omega^{(1,1)} & = & \mathbf{K}^3 \\
     \hline
   \end{array}
\end{equation}
where $\mathbf{k}^{x}$ is the triplet of antiself-dual $2$-forms on the base manifold associated with the triplet of complex
structures forming the quaternionic algebra and $\mathbf{K}^{x}$ is the analogous triplet of anti-self-dual $2$-forms on the
target manifold. On the base manifold we choose the complex coordinates according to the complex structure implicitly implied by
the choice (\ref{ipercallero}), namely we set:
\begin{equation}\label{basemancomp}
    \begin{array}{lclclcl}
       z^1 & = & \mathrm{Z} +{\rm i} \mathrm{U} & ; & z^2 & = & \mathrm{Y} - {\rm i} \mathrm{X} \\
       \bar{z}^{1^\star} & = & \mathrm{Z} -{\rm i} \mathrm{U} & ; & \bar{z}^{2^\star} & = & \mathrm{Y} + {\rm i} \mathrm{X}  \\
     \end{array}
\end{equation}
and we obtain:
\begin{equation}\label{gioadelcolle}
    \begin{array}{lcccr}
       \omega^{(2,0)} & = & -\, 2\,{\rm i} \, \mathrm{d}z^1 \wedge \mathrm{d}z^2 & = & -\,
       {\rm i} \, \epsilon_{\alpha\beta} \mathrm{d}z^{\alpha} \wedge \mathrm{d}z^{\beta} \\
     \omega^{(0,2)} & = & \, 2\,{\rm i} \, \mathrm{d}\bar{z}^{\bar{1}} \wedge \mathrm{d}\bar{z}^{\bar{2}} & = & \,
       {\rm i} \, \epsilon_{\bar{\alpha}\bar{\beta}} \mathrm{d}\bar{z}^{\bar{\alpha}} \wedge
       \mathrm{d}\bar{z}^{\bar{\beta}}\\
       \omega^{(1,1)} & = & {\rm i} \left( dz^{1} \wedge d\bar{z}^{\bar{1}} +
       dz^{2} \wedge d\bar{z}^{\bar{2}}\right) & = & {\rm i} \, \delta_{\alpha\bar{\beta}}
       \mathrm{d}{z}^{\alpha} \wedge \mathrm{d}\bar{z}^{\bar{\beta}}
     \end{array}
\end{equation}
Similarly let us introduce a set of complex coordinates on the target manifold well-adapted to the complex structure implicitly
implied by the choice (\ref{ipercallero}). We name them
\begin{equation}\label{ghinoditacco}
    q^j \quad ; \quad \bar{q}^{\bar{\jmath}} \quad ; \quad j \, = \, 1,2,\dots, 2m \quad \mbox{for } \mbox{dim} \, \mathcal{N} \,
    =  \, 4 \, m
\end{equation}
Using these coordinates we necessarily obtain:
\begin{equation}\label{disperacolle}
    \begin{array}{lcl}
       \Omega^{(2,0)} &  = &
       \Omega_{ij} \, \mathrm{d}q^{i} \wedge \mathrm{d}q^{j} \\
     \Omega^{(0,2)} & = & \overline{\Omega}_{\bar{\imath}\bar{\jmath}}  \,\mathrm{d}\bar{q}^{\bar{\imath}}
     \wedge \mathrm{d}\bar{q}^{\bar{\jmath}}\\
       \Omega^{(1,1)} &  = & {\rm i} \, g_{i\bar{\jmath}}\,
       \mathrm{d}{q}^{i} \wedge \mathrm{d}\bar{q}^{\bar{\jmath}}
     \end{array}
\end{equation}
where $g_{i\bar{\jmath}}$ is the HyperK\"ahler metric of the target manifold, which, in particular, is K\"ahler.
\par
Having fixed these conventions one easily retrieves the forms of the three complex structures arranged according to:
\begin{equation}\label{ipercallero}
   \begin{array}{|ccc|ccc|}
   \hline
     j^+ & = & \frac{j^{2}+{\rm i}j^1}{\sqrt{2}} &
     J^+ & = & \frac{J^{2}+{\rm i}J^1}{\sqrt{2}} \\
     j^- & = & \frac{j^{2}-{\rm i}j^1}{\sqrt{2}} &
     J^{-} & = & \frac{J^{2}-{\rm i}J^1}{\sqrt{2}} \\
     j^{0} & = & j^3 &
     J^0 & = & J^3 \\
     \hline
   \end{array}
\end{equation}
and one explicitly obtains:
\begin{equation}
\begin{array}{rclcrclcrclcrcl}
  \left(j^+\right)_\alpha^{\phantom{\alpha}\bar{\beta}} &=& \, - \, {\rm i} \, \epsilon_\alpha^{\phantom{\alpha}\bar{\beta}} &,&
  \left(j^+\right)_{\bar{\alpha}}^{\phantom{\alpha}{\beta}} &=& 0 &,&
\left(j^-\right)_{\bar{\alpha}}^{\phantom{\alpha}{\beta}} &=&  \, {\rm i} \,
  \epsilon_{\bar{\alpha}}^{\phantom{\alpha}{\beta}} &,&
\left(j^-\right)_{{\alpha}}^{\phantom{\alpha}{\bar{\beta}}} &=&  \, 0 \\
\left(j^0\right)_\alpha^{\phantom{\alpha}\bar{\beta}} &=& 0 &,&
  \left(j^0\right)_{\bar{\alpha}}^{\phantom{\alpha}{\beta}} &=& 0 &,&
\left(j^0\right)_{{\alpha}}^{\phantom{\alpha}{\beta}} &=&  \, {\rm i} \delta_{\alpha}^{\phantom{\alpha}\beta}&,&
\left(j^0\right)_{\bar{\alpha}}^{\phantom{\alpha}{\bar{\beta}}} &=& - \, {\rm i}
\delta_{\bar{\alpha}}^{\phantom{\alpha}\bar{\beta}} \\
\end{array} \label{tirabusson}
\end{equation}
for the base manifold and:
\begin{equation}
\begin{array}{rclcrclcrclcrcl}
  \left(J^+\right)_i^{\phantom{i}\bar{\jmath}} &=& \, - \, {\rm i} \, \Omega_i^{\phantom{i}\bar{\jmath}} &,&
  \left(J^+\right)_{\bar{\imath}}^{\phantom{i}{j}} &=& 0 &,&
\left(J^-\right)_{\bar{\imath}}^{\phantom{i}{j}} &=&  \, {\rm i} \,
  \bar{\Omega}_{\bar{\imath}}^{\phantom{i}{j}} &,&
\left(J^-\right)_{{i}}^{\phantom{i}{\bar{\jmath}}} &=&  \, 0 \\
\left(J^0\right)_i^{\phantom{i}\bar{\jmath}} &=& 0 &,&
  \left(J^0\right)_{\bar{\imath}}^{\phantom{i}{j}} &=& 0 &,&
\left(J^0\right)_{{i}}^{\phantom{i}{j}} &=&  \, {\rm i} \delta_{i}^{\phantom{i}j} &,&
\left(J^0\right)_{\bar{\imath}}^{\phantom{i}{\bar{\jmath}}} &=& - \, {\rm i} \delta_{\bar{\imath}}^{\phantom{i}\bar{\jmath}}
\end{array}
 \label{stoper}
\end{equation}
for the target manifold. In the above eq.(\ref{stoper}) we have defined:
\begin{equation}\label{deinisco}
    \Omega_i^{\phantom{i}\bar{\jmath}} \, = \, \Omega_{i\ell} \, g^{\ell\bar{\jmath}} \quad ; \quad
    \bar{\Omega}_{\bar{\imath}}^{\phantom{i}j}\, = \, \Omega_{\bar{\imath}\bar{\ell}} \, g^{\bar{\ell} j}
\end{equation}
Using this complex coordinate basis, the triholomorphicity conditions $\mathcal{E}_\mu^u \, = \,0$ defined by
eq.(\ref{triholequa}) become the following ones:
\begin{eqnarray}
  \partial_{\bar{\alpha}} \, q^i &=& 0 \label{holomorfo}\\
 \partial_\alpha \, q^i &=& \, - \,{\rm i} \, \ft 12 \, \epsilon_{\alpha}^{\phantom{\alpha}\bar{\beta}}\,
 \partial_{\bar{\beta}} \, \bar{q}^{\bar{\ell}} \, \bar{\Omega}_{\ell}^{\phantom{\ell}i}
 \label{triholomorfo}
\end{eqnarray}
plus the complex conjugates of the above.
\par
Equations (\ref{holomorfo},\ref{triholomorfo}) are quite revealing. The first of them (\ref{holomorfo}) tells us that any
triholomorphic map is in particular holomorphic. The second constraint relates the holomorphic derivatives of the holomorphic
coordinates to the anti-holomorphic derivatives of the anti-holomorphic coordinates through the unique $\Omega^{(2,0)}$ form of
the target manifold. From (\ref{holomorfo},\ref{triholomorfo}) another property is also immediately evident. If $q^i$ depends
holomorphically only from one of the two coordinates of the base manifold then the triholomorphicity conditions are automatically
satisfied. Hence any holomorphic embedding of a two dimensional subspace of the $4$-dimensional base manifold into the target
manifold is triholomorphic. This means that all instantons of $2D$ sigma-model are included in the space of hyperinstantons of
the 4D sigma-model. A quite relevant inclusion!
\par
Let us next consider the transcription of topological action (\ref{paronzo}) into complex formalism. Setting $\mathrm{O}\, = \,
1$ as already explained we obtain:
\begin{eqnarray}\label{topolinazione}
    \mathcal{A}_{top}[ q,t] & = & \int_{\mathcal{M}_4} \left(
    \omega^{(2,0)}\wedge q_\star\left[\Omega^{(0,2)}(t)\right]\, + \,
    \omega^{(0,2)}\wedge q_\star \left[\Omega^{(2,0)}(t)\right]
    \omega^{(1,1)}\wedge q_\star \left[\Omega^{(1,1)}(t)\right] \right)
\end{eqnarray}
where $t$ denote the $3\times n_+$ parameters of the manifold of HyperK\"ahler structure deformations:
\begin{equation}\label{basilisco}
   \mathcal{M}_{HK} \, = \, \frac{\mathrm{SO(3,n_+)}}{\mathrm{SO(3)\times SO(n_+)}} / \mathrm{O(3,n_+,\mathbb{Z})}
\end{equation}
$n_+$ being the number of self-dual $2$-forms which is $3$ for $\mathbb{R}^4$ and $19$ for K3. In equation (\ref{topolinazione})
we have emphasized that the HyperK\"ahler form depend on their moduli.
\par
Relying on eq.(\ref{kallerotti}) the topological action can be rewritten also in an equivalent way where the dependence of the
moduli-parameters becomes explicit:
\begin{eqnarray}
    \mathcal{A}_{top}[ q,t] & = & \mathbb{L}_{x}^{\phantom{x}J}(t) \times \int_{\mathcal{M}_4}
    \mathbf{k}^x \wedge  q_\star\left[\alpha_J\right] \label{pipponazione}
\end{eqnarray}
This way of rewriting the topological action is very useful in order to discuss the
right choice of the topological observables.

\section{About the coset $\mathrm{SO(3,3)/SO(3)\times SO(3)}$}
\label{solvabile}
The group $\mathrm{SO(3,3)}$ corresponds to the maximal non-compact real form of the Lie algebra $\mathrm{A_3} \sim \mathrm{D_3}$.
It follows that $\mathrm{SO(3,3)}$ is locally isomorphic to $\mathrm{SL(4,\mathbb{R})}$.
Indeed $\mathrm{SL(4,\mathbb{R})}$ is just the spinor representation of $\mathrm{SO(3,3)}$. Clearly this implies that:
\begin{equation}\label{isumurfu}
 \mathbb{H}_{3,3} \, \equiv\,    \frac{\mathrm{SO(3,3)}}{\mathrm{SO(3) \times SO(3)}} \, \simeq \,
 \frac{\mathrm{SL(4,\mathbb{R})}}{\mathrm{SO(4) }}
\end{equation}
Hence the coset manifold we are interested in can be alternatively viewed as the space of all  $3\times 3$ matrices
$\mathfrak{T}$,  or as the space of symmetric  $4\times 4$,  matrices $h$ with determinant one.
\subsection{The $\mathrm{SO(3,3)}\sim \mathrm{SL(4,R)}$ isomorphism, flat metrics and HyperK\"ahler structures}
\label{isomorfo} The aforementioned local isomorphism has a clearcut geometrical interpretation that is quite relevant in the
context of the topological field theory we consider. On one side, as we have seen in eq.s (\ref{intersecuto1},\ref{eta33primus})
and (\ref{cosettus},\ref{kallerotti}), the elements of the coset $\frac{\mathrm{SO(3,3)}}{\mathrm{SO(3) \times SO(3)}}$
parameterize the deformations of the flat HyperK\"ahler structures of $\mathbb{R}^4$. On the other hand the $4\times4$ symmetric
matrices with determinant one parameterize the space of flat constant metrics with unit volume. This observation provides the
means of geometrically constructing the isomorphism spelled out above.
\par
Let us consider the reference flat metric $h^{0}_{uv}\, = \, \delta_{uv}$ and the reference pair of quaternionic algebras of
tri-complex structures, provided, by the antiself-dual and antiself-dual 't Hooft matrices $J^x_0 \, = \, J^{-|x}\, ,
J^{\dot{x}}_0 \, = \, J^{+|\dot{x}} $:
\begin{eqnarray}\label{referenceJ}
    J_0^x \, J_0^y & = & - \, \delta^{xy} \, \mathbf{1} \, + \, \epsilon^{xyz} \,J_0^y \nonumber\\
     J_0^{\dot{x}} \, J_0^{\dot{y}} & =& - \, \delta^{\dot{x}\dot{y}} \, \mathbf{1} \, + \,
     \epsilon^{\dot{x}\dot{y}\dot{z}} \,J_0^{\dot{y}}
\end{eqnarray}
The indices of the complex structure are the first up, the second down, since, by definition, they are maps of the tangent bundle
into itself. From the pair of tri-complex structures we obtain the pair of HyperK\"ahler forms by lowering the first index with
the metric:
\begin{equation}\label{referenceK}
    K^{0|I}_{uv} \, = \, h^0_{us} \,\left(J_0^I\right)^{s}_{\phantom{s}v} \quad ; \quad \mathbf{K}^{0|I} \, = \,
    {K}^{0|I}_{uv} \, \mathrm{d}q^u\wedge \mathrm{d}q^v \quad \quad I=\{x,\dot{x}\}
\end{equation}
Let us next consider  a generic constant flat metric $h_{uv}$ with the same signature and determinant one. It can be written as
follows:
\begin{equation}\label{metruzzapiatta}
    h \, = \, S^T \, h_0\, S \quad ; \quad S\, \in \, \mathrm{SL(4,\mathbb{R})}
\end{equation}
Clearly $S$ and $S^\prime \, = \,S\, O$ with $O \in \mathrm{SO(4)}$ give rise to the same flat metric, which shows that
$\frac{\mathrm{SL(4,\mathbb{R})}}{\mathrm{SO(4) }}$ is the space of all metrics $h_{uv}$ on $\mathbb{R}^4$. The writing
(\ref{metruzzapiatta}) leads to conclude that:
\begin{equation}\label{cocconulla}
    J^I \, = \, S^{-1} \, J^I_0 \, S
\end{equation}
is the new pair  of tri-complex structures satisfying the same quaternionic algebras:
\begin{eqnarray}\label{referenceJ}
    J^x \, J^y & = & - \, \delta^{xy} \, \mathbf{1} \, + \, \epsilon^{xyz} \,J^y \nonumber\\
    J^{\dot{x}} \, J^{\dot{y}} & = & - \, \delta^{\dot{x}\dot{y}} \, \mathbf{1} \, + \, \epsilon^{\dot{x}\dot{y}\dot{z}} \,J^{\dot{y}}
\end{eqnarray}
on $\mathbb{R}^4$ endowed with the metric $h_{uv}$. Lowering the first index of the new pair of tri-complex structures with the
metric $h$ we obtain the new pair of  HyperK\"ahler forms:
\begin{equation}\label{forgetto}
    K^I \, = \, S^T \, K^{0|I} \, S \quad  ; \quad \mathbf{K}^{I} \, = \,
    {K}^{I}_{uv} \, \mathrm{d}q^u\wedge \mathrm{d}q^v
\end{equation}
Next calculating the intersection product of the new pair of HyperK\"ahler forms, we obtain:
\begin{equation}\label{colluma}
    \mathbf{K}^{I}\wedge  \mathbf{K}^{J} \, = \, \left(\mbox{det} S\right)\times\mathbf{K}_0^{I}\wedge  \mathbf{K}_0^{J}
    \, = \, \left(\mbox{det} S\right) \, 8 \, \times \, \eta_{IJ} \, \mbox{Vol}
\end{equation}
Hence as long as the determinant of the matrix $S$ is one, the intersection matrix of the $K$-forms is preserved. Since
${K}^{I}_{uv}$ are anyhow antisymmetric $4\times 4$ matrices, they can be linearly expanded on the basis of such matrices
provided by the 't Hooft ones. This yields
\begin{equation}\label{isomorfina}
    K^I \, = \, L^{I}_{\phantom{I}J} \, K^{0|J} \, = \, S^T \, K^{0|I} \, S \quad ; \quad L \, \in \, \mathrm{SO(3,3)}
\end{equation}
which is the explicit form of the local isomorphism and provides also the algorithm to construct explicitly $L$ starting from
$S$. The locality of the isomorphism is clear from the fact that $S$ and $-\, S$ yield the same $L$.
\subsection{The solvable parameterization and fractional linear transformations}
\label{fractional}
 It is useful to recall that the group
$\mathrm{SO(\bar{n},\bar{n})}$ works also as the group of electric-magnetic duality rotation on a set of $n$ field strengths that
are $(2p+1)$-forms in dimensions $D=4p+2$ just as the symplectic group $\mathrm{Sp(\bar{n},\bar{n})}$ does the same job on $n$
field strengths that are $2p$-forms in dimensions $D=4p$ \cite{mieletture}. This fact provides the proper setup to parameterize
the coset manifold (\ref{isumurfu}) in terms of a $3\times 3$ matrix with nice fractional linear transformation under
$\mathrm{SO(3,3)}$. Suppose we had some theory of gauge $2$-forms in six dimensions and that the number of such gauge forms were
precisely three. Then, calling $\mathbf{F}_I^{[3]}$ the corresponding $3$-form field strengths we would have the following
lagrangian:
\begin{equation}\label{combinat}
    \mathcal{L}\, = \, \gamma_{IJ}(\phi) \, \mathbf{F}^I_{[3]}\wedge \star_g \mathbf{F}^J_{[3]}\, + \,
    \theta_{IJ}(\phi) \, \mathbf{F}^I_{[3]}\wedge  \mathbf{F}^J_{[3]}\,
\end{equation}
where $\star_g$ denotes the Hodge dual with respect to the space-time metric and $\gamma_{IJ}(\phi)$
and $\theta(\phi)$ are two $3\times 3$ matrices respectively symmetric and antisymmetric that depend
on the scalar fields included in the theory.  As shown in
\cite{mieletture} the most general electric-magnetic duality rotations  mixing field equations and
Bianchi identities that are admitted by the above theory are described  in the following way. Define the matrix:
\begin{equation}\label{turiesko}
    \mathfrak{T}(\phi) \, \equiv \, \gamma_{IJ}(\phi) \, + \, \theta_{IJ}(\phi)
\end{equation}
and consider a generic $\mathrm{SO(3,3)}$-matrix:
\begin{equation}\label{cospirato}
\left(\begin{array}{c|c}
        A & B \\
        \hline
        C & D
      \end{array}
\right)    \,\equiv \, \Lambda_{dual} \, \in \, \mathrm{SO(3,3)}\quad \Leftrightarrow \quad \Lambda_{dual}^T
\left(\begin{array}{c|c}
        0 & \mathbf{1}_{3\times 3} \\
        \hline
         \mathbf{1}_{3\times 3} &0
      \end{array}
\right)\, \Lambda_{dual} \, = \, \underbrace{\left(\begin{array}{c|c}
        0 & \mathbf{1}_{3\times 3} \\
        \hline
         \mathbf{1}_{3\times 3} &0
      \end{array}
\right)\,}_{\eta_{dual}}
\end{equation}
then the most general electric-magnetic duality rotation is represented on the matrix $ \mathfrak{T}(\phi)$ by the following linear fractional transformation:
\begin{equation}\label{consonante}
    \gamma_{IJ}^\prime \, + \, \theta_{IJ}^\prime \, \equiv \, \mathfrak{T}^\prime \, =
    \, \left(A \, \mathfrak{T} \,+ \,B\right) \, \left( C \, \mathfrak{T} \, + \, D\right )^{-1}
\end{equation}
For this reason in this section we consider three alternative forms of the $\so(3,3)$ invariant metric, namely:
 \begin{eqnarray}\label{etadiag}
    \eta_{solv}&= &\left(
\begin{array}{llllll}
 0 & 0 & 0 & 0 & 0 & 1 \\
 0 & 0 & 0 & 0 & 1 & 0 \\
 0 & 0 & 0 & 1 & 0 & 0 \\
 0 & 0 & 1 & 0 & 0 & 0 \\
 0 & 1 & 0 & 0 & 0 & 0 \\
 1 & 0 & 0 & 0 & 0 & 0
\end{array}
\right) \quad ; \quad \eta_{diag}\, = \,\left(
\begin{array}{llllll}
 -1 & 0 & 0 & 0 & 0 & 0 \\
 0 & -1 & 0 & 0 & 0 & 0 \\
 0 & 0 & -1 & 0 & 0 & 0 \\
 0 & 0 & 0 & 1 & 0 & 0 \\
 0 & 0 & 0 & 0 & 1 & 0 \\
 0 & 0 & 0 & 0 & 0 & 1
\end{array}
\right)\nonumber\\
\eta_{dual}& = & \,\left(
\begin{array}{llllll}
 0 & 0 & 0 & 1 & 0 & 0 \\
 0 & 0 & 0 & 0 & 1 & 0 \\
 0 & 0 & 0 & 0 & 0 & 1 \\
 1 & 0 & 0 & 0 & 0 & 0 \\
 0 & 1 & 0 & 0 & 0 & 0 \\
 0 & 0 & 1 & 0 & 0 & 0
\end{array}
\right)
 \end{eqnarray}
They are related one to the other by the following transformations:
\begin{eqnarray}\label{trasformazia}
    \eta_{diag} & =& \mathrm{M}^T \,  \eta_{solv} \, \mathrm{M}\quad ; \quad \mathrm{M}\, = \, \left(
\begin{array}{llllll}
 -\frac{1}{\sqrt{2}} & 0 & 0 &
   \frac{1}{\sqrt{2}} & 0 & 0 \\
 0 & -\frac{1}{\sqrt{2}} & 0 & 0 &
   \frac{1}{\sqrt{2}} & 0 \\
 0 & 0 & -\frac{1}{\sqrt{2}} & 0 & 0 &
   \frac{1}{\sqrt{2}} \\
 0 & 0 & \frac{1}{\sqrt{2}} & 0 & 0 &
   \frac{1}{\sqrt{2}} \\
 0 & \frac{1}{\sqrt{2}} & 0 & 0 &
   \frac{1}{\sqrt{2}} & 0 \\
 \frac{1}{\sqrt{2}} & 0 & 0 &
   \frac{1}{\sqrt{2}} & 0 & 0
\end{array}
\right)\nonumber\\
\eta_{dual} & = & \mathrm{N}^T \,  \eta_{diag} \, \mathrm{N}\quad ; \quad \mathrm{N}\, = \, \frac{1}{\sqrt{2}}\,\left(
\begin{array}{llllll}
 -1 & 0 & 0 & 1 & 0 & 0 \\
 0 & -1& 0 & 0 & 1 & 0 \\
 0 & 0 & -1 & 0 & 0 & 1 \\
 1 & 0 & 0 & 1 & 0 & 0 \\
 0 & 1& 0 & 0 & 1 & 0 \\
 0 & 0 & 1 & 0 & 0 & 1
\end{array}
\right)
\end{eqnarray}
Correspondingly the group elements of  $\mathrm{SO(3,3)}$ are alternatively defined by one of the three
quadratic constraints:
\begin{eqnarray}\label{costrettiq}
\Lambda \in \mathrm{SO(3,3)} & \Leftrightarrow &  \Lambda_{solv}^T \,\eta_{solv} \, \Lambda_{solv} \, = \,
\eta_{solv} \quad; \quad \Lambda_{diag}^T \,\eta_{diag} \,
    \Lambda_{diag} \, = \, \eta_{diag}\nonumber\\
     &&  \Lambda_{dual}^T \,\eta_{dual} \,\Lambda_{dual} \, = \, \eta_{dual}
\end{eqnarray}
and the relation between the explicit form of the same group element in the three bases is given below:
\begin{equation}\label{basarelazia}
    \Lambda_{diag} \, = \, \mathrm{M}^{-1} \, \Lambda_{solv} \, \mathrm{M} \quad ; \quad
    \Lambda_{dual} \, = \, \mathrm{N}^{-1} \, \Lambda_{diag} \, \mathrm{N}
\end{equation}
Recalling eq.s (\ref{intersecuto1},\ref{eta33primus}) we see that the basis where the $\so(3,3)$ metric is diagonal is the
natural one for the cohomology lattice since $  8 \times \eta_{diag}$ is the  intersection form of $\mathbb{R}^4$. On the other
hand the solvable basis where the $\so(3,3)$ metric has an anti-diagonal form is the best suited to derive the solvable
parametrization of the non compact coset under consideration. Indeed, according to a well established mathematical theory
\cite{solvabli}, a maximally non-compact coset such as ours is metrically equivalent to a solvable group-manifold, in particular
to the exponential map of the Borel subalgebra. In the solvable basis where the invariant metric is given by $\eta_{solv}$, the
Borel subalgebra $\mathfrak{B}\left[\so(3,3)\right]$ is composed by upper triangular matrices and its exponentiation is
particularly simple.
\par
The most general element of the Borel subalgebra in the solvable basis is the following one:
\begin{equation}\label{borellianusA}
    \mathbf{B}_a \, = \, \sum_i^9 \, t_i \, \mathrm{B}^i_a \, = \, \left(
\begin{array}{llllll}
 h_1 & \tau _1 & \tau _3 & \tau _5 & \tau _6
   & 0 \\
 0 & h_2 & \tau _2 & \tau _4 & 0 & -\tau _6
   \\
 0 & 0 & h_3 & 0 & -\tau _4 & -\tau _5 \\
 0 & 0 & 0 & -h_3 & -\tau _2 & -\tau _3 \\
 0 & 0 & 0 & 0 & -h_2 & -\tau _1 \\
 0 & 0 & 0 & 0 & 0 & -h_1
\end{array}
\right)
\end{equation}
where we have introduced the set of 9 parameters:
\begin{equation}\label{parametriBorello}
  t_i \, = \,  \left\{h_1,h_2,h_3,\tau _1,\tau _2,\tau
   _3,\tau _4,\tau _5,\tau _6\right\}
\end{equation}
the generator $\mathrm{B}^i_a$ being the matrix coefficient of the corresponding parameter in the expansion of
$\mathbf{B}_a$. The first three parameters $h_{1,2,3}$ correspond to the Cartan generators while the remaining six
parameters $\tau_i$ correspond to the six positive roots of the $\so(3,3)$ Lie algebra.
\par
Next we can introduce the coset representative in the solvable basis by setting:
\begin{equation}\label{matusatto}
    \mathbb{L}_a(t)\, = \, \prod_{i}^9 \, \exp\left[t_i \mathrm{B}_a^i\right] \, = \,\exp\left[t_1
    \mathrm{B}_a^1\right] \dots  \exp\left[t_9
    \mathrm{B}_a^9\right]
\end{equation}
and the explicit result is the following one:
\begin{eqnarray}\label{Lgrasso}
   & \mathbb{L}_{solv}(t)\, = \,& \nonumber\\
   &\left(
\begin{array}{llllll}
 e^{h_1} & e^{h_1} \tau _1 & e^{h_1}
   \left(\tau _1 \tau _2+\tau _3\right) &
   e^{h_1} \left(\tau _1 \tau _4+\tau
   _5\right) & e^{h_1} \left(\tau
   _6-\left(\tau _1 \tau _2+\tau _3\right)
   \tau _4\right) & -e^{h_1} \left(\tau _3
   \tau _5+\tau _1 \left(\tau _2 \tau
   _5+\tau _6\right)\right) \\
 0 & e^{h_2} & e^{h_2} \tau _2 & e^{h_2}
   \tau _4 & -e^{h_2} \tau _2 \tau _4 &
   -e^{h_2} \left(\tau _2 \tau _5+\tau
   _6\right) \\
 0 & 0 & e^{h_3} & 0 & -e^{h_3} \tau _4 &
   -e^{h_3} \tau _5 \\
 0 & 0 & 0 & e^{-h_3} & -e^{-h_3} \tau _2 &
   -e^{-h_3} \tau _3 \\
 0 & 0 & 0 & 0 & e^{-h_2} & -e^{-h_2} \tau
   _1 \\
 0 & 0 & 0 & 0 & 0 & e^{-h_1}
\end{array}
\right)&\nonumber\\
\end{eqnarray}
Transformed to the dual basis the coset representative $\mathbb{L}_{solv}(t)$ takes the following form:
\begin{eqnarray}
 &\mathbb{L}_{dual}(t) = & \nonumber \\
  &\left(
\begin{array}{lll|lll}
 e^{h_1} & e^{h_1} \tau _1 & e^{h_1}
   \left(\tau _1 \tau _2+\tau _3\right) &
   -e^{h_1} \left(\tau _3 \tau _5+\tau _1
   \left(\tau _2 \tau _5+\tau
   _6\right)\right) & e^{h_1} \left(\tau
   _6-\left(\tau _1 \tau _2+\tau _3\right)
   \tau _4\right) & e^{h_1} \left(\tau _1
   \tau _4+\tau _5\right) \\
 0 & e^{h_2} & e^{h_2} \tau _2 & -e^{h_2}
   \left(\tau _2 \tau _5+\tau _6\right) &
   -e^{h_2} \tau _2 \tau _4 & e^{h_2} \tau
   _4 \\
 0 & 0 & e^{h_3} & -e^{h_3} \tau _5 &
   -e^{h_3} \tau _4 & 0 \\
   \hline
 0 & 0 & 0 & e^{-h_1} & 0 & 0 \\
 0 & 0 & 0 & -e^{-h_2} \tau _1 & e^{-h_2} &
   0 \\
 0 & 0 & 0 & -e^{-h_3} \tau _3 & -e^{-h_3}
   \tau _2 & e^{-h_3}
\end{array}
\right)& \nonumber\\
\label{bangio}
\end{eqnarray}
where we easily recognize the blocks $A(t)$,$B(t)$,$C(t)$ and $D(t)$. Given this form we can reconstruct the
parameterization of the coset by means of a matrix $\mathfrak{T}$ regarding it as the fractional linear
transform of the identity matrix, namely setting:
\begin{equation}\label{golirdino}
    \mathfrak{T}(t) \, = \, \left(A(t) \, \mathbf{1}_{3\times 3} \, + \, B\right) \, \left( C(t) \,
    \mathbf{1}_{3\times 3} \, + \, D(t)\right)^{-1}
    \end{equation}
The result is encoded in the following:
{\scriptsize
\begin{eqnarray}\label{gothT}
   & \mathfrak{T}(t) \, = &\nonumber\\
   &\left(
\begin{array}{lll}
 e^{2 h_1} \left(\left(\tau _2^2+1\right)
   \tau _1^2+2 \tau _2 \tau _3 \tau
   _1+\tau _3^2+1\right) & e^{h_1+h_2}
   \left(\tau _1 \left(\tau
   _2^2+1\right)-\tau _3 \tau _4+\tau _2
   \left(\tau _3+\tau _5\right)+\tau
   _6\right) & e^{h_1+h_3} \left(\tau
   _3+\tau _1 \left(\tau _2+\tau
   _4\right)+\tau _5\right) \\
 e^{h_1+h_2} \left(\tau _1 \left(\tau
   _2^2+1\right)+\tau _3 \left(\tau
   _2+\tau _4\right)-\tau _2 \tau _5-\tau
   _6\right) & e^{2 h_2} \left(\tau
   _2^2+1\right) & e^{h_2+h_3} \left(\tau
   _2+\tau _4\right) \\
 e^{h_1+h_3} \left(\tau _3+\tau _1
   \left(\tau _2-\tau _4\right)-\tau
   _5\right) & e^{h_2+h_3} \left(\tau
   _2-\tau _4\right) & e^{2 h_3}
\end{array}
\right)&\nonumber\\
\end{eqnarray}
}
which provides the relation between the solvable parameterization of the coset and its projective parameterization in terms of a matrix
$\mathfrak{T}$. What we learn from this example is that in what we called the dual basis a coset representative that
transforms the unit matrix into a generic $\mathfrak{T}$ can be chosen upper triangular. Indeed with reference to eq.(\ref{turiesko})
we can set:
\begin{equation}\label{canopello}
    \mathbb{L}(\mathfrak{T}) \, \in \, \mathrm{SO(3,3)} \quad ; \quad  \mathbb{L}(\mathfrak{T})\, = \, \left( \begin{array}{c|c}
                                                \gamma^{\ft12} &\theta \,  \gamma^{-\ft12} \\
                                                \hline
                                                0& \gamma^{-\ft12}
                                              \end{array}
    \right)
\end{equation}
\subsection{The off-diagonal parameterization}
\label{offodiag} Another very elegant and symmetric parametrization that has just the inconvenience of involving  non polynomial
matrix functions is the classical off-diagonal one. In the diagonal basis the $\mathbb{H}_{3,3}$ coset representative can be
written as:
\begin{equation}\label{posaidone}
\mathbb{L}_{diag}(\Psi) \, \left(\begin{array}{c|c}
                                   \sqrt{1+\Psi \, \Psi^T} & \Psi \\
                                   \hline
                                   \Psi^T & \sqrt{1+\Psi^T \, \Psi}
                                 \end{array}
\right)
\end{equation}
where $\Psi$ is a generic $3\times 3$ matrix that encodes all the 9 parameters of the coset.
\begin{landscape}
\section{Some large formulas}
 {\tiny
\begin{equation}\label{fanciullus}
  \mathcal{Q}_{\mathcal{IJ}} \, = \,   \left(
\begin{array}{cccccccccccc}
 \gamma _{3,3}-\psi _{3,3} & 0 & 0 & 0 & 0 & 0 & 0 & 0 & -\psi _{3,2} & 0 &
   0 & \psi _{3,1} \\
 0 & \gamma _{2,2}+\psi _{2,2} & 0 & 0 & 0 & 0 & 0 & -\psi _{2,3} & 0 & 0 &
   -\psi _{2,1} & 0 \\
 0 & 0 & \gamma _{3,3}-\psi _{3,3} & 0 & 0 & 0 & 0 & 0 & \psi _{3,1} & 0 & 0
   & \psi _{3,2} \\
 0 & 0 & 0 & \gamma _{1,1}-\psi _{1,1} & 0 & 0 & -\psi _{1,3} & 0 & 0 & \psi
   _{1,2} & 0 & 0 \\
 0 & 0 & 0 & 0 & \gamma _{2,2}+\psi _{2,2} & 0 & 0 & \psi _{2,1} & 0 & 0 &
   -\psi _{2,3} & 0 \\
 0 & 0 & 0 & 0 & 0 & \gamma _{1,1}-\psi _{1,1} & \psi _{1,2} & 0 & 0 & \psi
   _{1,3} & 0 & 0 \\
 0 & 0 & 0 & -\psi _{1,3} & 0 & \psi _{1,2} & \gamma _{1,1}+\psi _{1,1} & 0
   & 0 & 0 & 0 & 0 \\
 0 & -\psi _{2,3} & 0 & 0 & \psi _{2,1} & 0 & 0 & \gamma _{2,2}-\psi _{2,2}
   & 0 & 0 & 0 & 0 \\
 -\psi _{3,2} & 0 & \psi _{3,1} & 0 & 0 & 0 & 0 & 0 & \gamma _{3,3}+\psi
   _{3,3} & 0 & 0 & 0 \\
 0 & 0 & 0 & \psi _{1,2} & 0 & \psi _{1,3} & 0 & 0 & 0 & \gamma _{1,1}+\psi
   _{1,1} & 0 & 0 \\
 0 & -\psi _{2,1} & 0 & 0 & -\psi _{2,3} & 0 & 0 & 0 & 0 & 0 & \gamma
   _{2,2}-\psi _{2,2} & 0 \\
 \psi _{3,1} & 0 & \psi _{3,2} & 0 & 0 & 0 & 0 & 0 & 0 & 0 & 0 & \gamma
   _{3,3}+\psi _{3,3} \\
\end{array}
\right)
\end{equation}}
\end{landscape}


\newpage
\begin{thebibliography}{99}
\bibitem{Fre:2015xaa}
  P.~Fre and A.~S.~Sorin,
  \emph{2-branes with Arnold-Beltrami Fluxes from Minimal D=7 Supergravity},
  arXiv:1504.06802 [hep-th].
\bibitem{PvNT} P.K. Townsend and P. van Nieuwenhuizen \emph{Gauged Seven Dimensional Supergravity} Phys. Lett.
\textbf{B125} (1983), 41
\bibitem{SalamSezgin} Abdus Salm and E. Sezgin \emph{SO(4) Gauging of $\mathcal{N}=2$ supergravity in seven dimensions}
Phys. Lett. \textbf{B126} (1983), 295
\bibitem{bershoffo1} E.Bergshoeff, I.G. Koh and E. Sezgin \emph{Yang-Mills Einstein supergravity in seven dimensions}
Phys. Rev. D, \textbf{32}, 6, (1985) 1353.
%
\bibitem{beltramus} E. Beltrami,  Opere matematiche, {\bf 4} (1889) 304.
%
\bibitem{Fre:2015mla}
  P.~Fre and A.~S.~Sorin,
  \emph{Classification of Arnold-Beltrami Flows and their Hidden Symmetries},
  arXiv:1501.04604 [math-ph].
%
\bibitem{arnoldorussopapero} V. I. Arnold, \emph{On the evolution of a magnetic feld under the action of transport
and diffusion}, in  \emph{Vladimir I. Arnold: Collected Works, Volume II, Hydrodynamics, Bifurcation Theory, and
Algebraic Geometry 1965-1972} (Edited by Alexander B. Givental, Boris A. Khesin, Alexander N. Varchenko, Victor A.
Vassiliev, Oleg Ya. Viro), 405 - 419, Springer-Verlag Berlin Heidelberg 2014. Originally published in: \emph{Some
Problems in Modern Analysis}, 8-21 (Russian), © Izd. MGU, Moscow 1984.
\bibitem{balubbo}
The ABC flows have been discovered by Gromeka in 1881, rediscovered by Beltrami \cite{beltramus}, and proposed for
study in the present context in \cite{arnoldorussopapero} and \cite{Childress}.
%
%
\bibitem{Childress}
S. Childress, \emph{Construction of steady-state hydromagnetic dynamos. I. Spatially periodic fields}, Report MF-53,
Courant Inst. of Math. Sci. (1967); \emph{New solutions of the kinematic dynamo problem}, J. Math. Phys.  {\bf 11}
(1970) 3063 - 3076.
\bibitem{Anselmi:1993wm}
  D.~Anselmi and P.~Fre,
  \emph{Topological sigma models in four-dimensions and triholomorphic maps},
  Nucl.\ Phys.\ B {\bf 416} (1994) 255
  [hep-th/9306080].
\bibitem{Andrianopoli:1996cm}
  L.~Andrianopoli, M.~Bertolini, A.~Ceresole, R.~D'Auria, S.~Ferrara, P.~Fre and T.~Magri,
  \emph{`N=2 supergravity and N=2 superYang-Mills theory on general scalar manifolds: Symplectic covariance, gaugings and the momentum map,}  J.\ Geom.\ Phys.\  {\bf 23} (1997) 111
  [hep-th/9605032].
\bibitem{Anselmi:1992tj}
  D.~Anselmi and P.~Fre,
  \emph{Twisted N=2 supergravity as topological gravity in four-dimensions},
  Nucl.\ Phys.\ B {\bf 392} (1993) 401
  [hep-th/9208029].
\bibitem{Anselmi:1992tz}
  D.~Anselmi and P.~Fre,
  \emph{Topological twist in four-dimensions, R duality and hyperinstantons},
  Nucl.\ Phys.\ B {\bf 404} (1993) 288
  [hep-th/9211121].
\bibitem{Anselmi:1994bu}
  D.~Anselmi and P.~Fre,
  \emph{Gauged hyper - instantons and monopole equations},
  Phys.\ Lett.\ B {\bf 347} (1995) 247
  [hep-th/9411205].
\bibitem{Fayet:1976cr}
  P.~Fayet and S.~Ferrara,
  \emph{Supersymmetry},
  Phys.\ Rept.\  {\bf 32} (1977) 249.
\bibitem{n2wonder} P.~Fre and P.~Soriani,
  \emph{The N=2 wonderland: From Calabi-Yau manifolds to topological field theories},
  Singapore, Singapore: World Scientific (1995) 468 p
%
\bibitem{solvabli}
  P.~Fre and A.~S.~Sorin,
  \emph{Supergravity Black Holes and Billiards and Liouville integrable structure of dual Borel algebras}
  JHEP {\bf 1003} (2010) 066
  [arXiv:0903.2559 [hep-th]].\\
P.~Fre, V.~Gili, F.~Gargiulo, A.~S.~Sorin, K.~Rulik and M.~Trigiante,
  \emph{Cosmological backgrounds of superstring theory and solvable algebras: Oxidation and branes}
  Nucl.\ Phys.\ B {\bf 685} (2004) 3
  [hep-th/0309237].\\
   L.~Andrianopoli, R.~D'Auria, S.~Ferrara, P.~Fre and M.~Trigiante,
  \emph{E(7)(7) duality, BPS black hole evolution and fixed scalars}
  Nucl.\ Phys.\ B {\bf 509} (1998) 463
  [hep-th/9707087].\\
  L.~Andrianopoli, R.~D'Auria, S.~Ferrara, P.~Fre, R.~Minasian and M.~Trigiante,
  \emph{Solvable Lie algebras in type IIA, type IIB and M theories,}
  Nucl.\ Phys.\ B {\bf 493} (1997) 249
  [hep-th/9612202].\\
  L.~Andrianopoli, R.~D'Auria, S.~Ferrara, P.~Fre and M.~Trigiante,
  \emph{RR scalars, U duality and solvable Lie algebras,}
  Nucl.\ Phys.\ B {\bf 496} (1997) 617
  [hep-th/9611014].
%
\bibitem{mieletture}
  P.~Fre,
  \emph{Lectures on special Kahler geometry and electric - magnetic duality rotations,}
  Nucl.\ Phys.\ Proc.\ Suppl.\  {\bf 45BC} (1996) 59
  [hep-th/9512043].\\
  P.~Fre,
\emph{Gaugings and other supergravity tools of p-brane physics},
  hep-th/0102114.




\end{thebibliography}
\end{document}